\documentclass[peerreview, draftclsnofoot, onecolumn, 12pt]{IEEEtran}		

\ifx\pdfoutput\undefined
\usepackage{graphicx}  
\else
\usepackage[pdftex]{graphicx}
\fi
				\usepackage{enumitem}	
				\usepackage{url}
				\usepackage[numbers,sort&compress]{natbib}
				\usepackage{array}							
				\usepackage{hyperref} 				
				\usepackage{stfloats}
				\usepackage{amssymb} 
				\usepackage{booktabs}
				\usepackage{multirow} 
				\usepackage{amsmath} 
				\usepackage{amsfonts}
				\usepackage[ansinew]{inputenc} % Get tildes
				\usepackage[caption=false,font=footnotesize]{subfig}
				\usepackage{amsthm} 
				\usepackage{microtype}
				\usepackage[table]{xcolor}
				\usepackage{algorithmic} 
				\usepackage[ruled,linesnumbered]{algorithm2e}	
				\usepackage{mathtools}

			\newlength{\TABLEUP}

\begin{document}

		% -- Paper and authors info ---------------------------------
			\title{Spatial Mappings for Planning\\ and Optimization of Cellular Networks}

			\author{David~González~G.,~\IEEEmembership{Member,~IEEE,} Harri Hakula, Antti Rasila, and~Jyri~Hämäläinen,~\IEEEmembership{Member,~IEEE}. %\IEEEmembership{Fellow,~OSA,}%\IEEEmembership{Life~Fellow,~IEEE}% <-this % stops a space
			\IEEEcompsocitemizethanks{
			\IEEEcompsocthanksitem 
					David~González~G. and Jyri Hämäläinen are with the Department of Communications and Networking, Aalto University, Finland. Corresponding 
					email: david.gonzalez.g@ieee.org. Harri Hakula and Antti Rasila are with the Department of Mathematics and Systems Analysis, Aalto University, Finland.
					}}
			\maketitle
			
			\vspace{-0.05cm}
		% -- Abstract ---------------------------------
			\begin{abstract} 
			In cellular networks, users are grouped into different cells and served by different access points (base stations) that provide wireless access to services and applications. In general, the service demand is very heterogeneous, non-uniformly distributed, and dynamic. Consequently, radio access networks create very \textit{irregular} topologies with more access points where service demand is concentrated. While this dynamism requires networks with the ability to adapt to time-varying conditions, the non-uniformity of the service demand makes the planning, analysis, and optimization difficult. In order to help with these tasks, a framework based on canonical domains and spatial mappings (e.g., conformal mapping) have recently been proposed. The idea is to carry out part of the planning in a canonical (perfectly symmetric) domain that is connected to the physical one (real-scenario) by means of a spatial transformation designed to map the access points \textit{consistently} with the service demand. This paper continues the research in that direction by introducing additional tools and possibilities to that framework, namely the use of centroidal Voronoi algorithms and non-conformal composite mappings. Moreover, power optimization is also introduced to the framework. The results show the usability and effectiveness of the proposed method and its promising research perspectives.
			\end{abstract}

\vspace{-0.05cm}
		% -- Keywords ---------------------------------
			\begin{IEEEkeywords}
					Cellular Networks, Network Planning, Conformal Mapping, Power Optimization, Voronoi \mbox{Tessellations}.
			\end{IEEEkeywords}
			%\IEEEpeerreviewmaketitle

		% -- SECTION: INTRODUCTION ---------------------------------
		
		\section{Introduction}\label{Sec:Intro}
\subsection{Context and motivation}\label{Sec:Intro:Context}

Radio access planning and optimization are fundamental tasks in cellular networks. Broadly speaking, planning refers to the tasks of determining the number, location, and configuration of access points to provide wireless access to users (and \textit{things})
to services and applications, with a certain targeted Quality of Service~(QoS). In particular, the problem of finding the number of access points is also referred to as \textsl{dimensioning}~\cite{7368887}, and this initial step aims at providing the required capacity for the service demand volume that is expected. However, in practice, both dimensioning and sites positioning are very difficult problems because the service demand is not uniformly distributed and it is quite diverse and dynamic. Nowadays, taking into account the continuous evolution of radio access technologies, and the new concepts and paradigms that are expected for the fifth generation~(5G) of cellular networks, the boundary between planning and optimization tasks becomes blurred. Indeed, according to the excellent work presented in~\cite{7368887}, planning and optimization are iterative tasks that go \textit{hand-in-hand}. In this line of thinking, the authors of~\cite{7462493} also pose the need for re-thinking planning. They emphasize the importance of distributing the service demand as evenly as possible among cells as a key criterion to achieve effective planning; a goal that in the opinion of the authors of~\cite{7462493} (and in our's) is a very valid way to enhance system performance. 

\begin{figure}[t]
	    		\centering	    	
	    		\includegraphics[width = 0.90\textwidth]{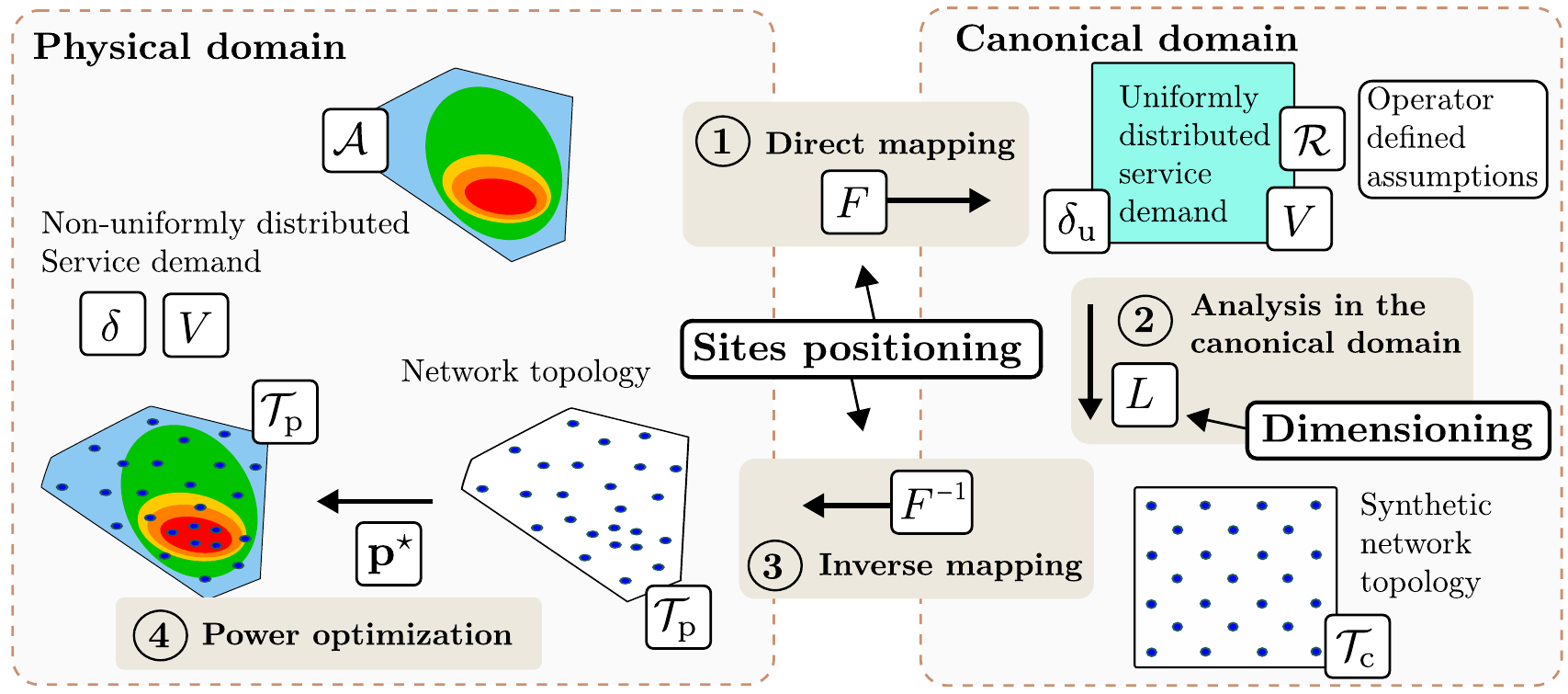}
	    		\vspace{-0.65cm}\caption{Planning and optimization based on canonical domains and spatial mappings. The original problem introduced in~\cite{7399421} comprised Steps~1, 2, and 3 (direct and inverse mapping and analysis in the canonical domain). The proposed enhanced framework includes: generalizations for Step~1, alternative methods for Step~3, and integration of Step~4 (power optimization).}
	    		\label{Fig:PROP_FRAMEWORK}    		
			\end{figure}

In our previous work~\cite{7399421}, also motivated by the aforementioned ideas, a novel framework for planning and optimization based on the use of \textsl{canonical domains} and spatial transformations was presented. Therein, planning is addressed by breaking the problem into two parts: dimensioning and sites positioning, as it is shown in Fig.~\ref{Fig:PROP_FRAMEWORK}. The central idea is to carry out the dimensioning in a dual/canonical domain in which the service demand is uniformly distributed. In this manner, a regular network topology ($\mathcal{T}_{\rm c}$) with the required number of access points is obtained (Step 2). Then, sites positioning is performed by mapping the regular network topology ($\mathcal{T}_{\rm c}$) from the canonical domain ($\mathcal{R}$) onto the physical domain ($\mathcal{A}$) by means of a spatial mapping $F^{-1}:\mathcal{R}\rightarrow\mathcal{A}$ that corresponds to the inverse of another (previously computed) mapping $F:\mathcal{A}\rightarrow\mathcal{R}$. The idea is that, if the mapping $F$ is designed (Step 1) such that it maps $\mathcal{A}$ onto $\mathcal{R}$ redistributing the service demand uniformly, then $F^{-1}$ will map (Step 3) the access points from $\mathcal{R}$ onto $\mathcal{A}$ in a compatible manner with the service demand, i.e., the network topology $\mathcal{T}_{\rm p}$. In~\cite{7399421}, conformal mapping~\cite{papamichael2010numerical}, a mature field in Complex Analysis~\cite{ahlfors1953complex}, but previously unheard of in the context of cellular networks, was proposed to obtain the mappings $F$ and $F^{-1}$.
\subsection{Contribution}\label{Sec:Intro:Contrib}
The idea introduced in~\cite{7399421} is new and it represents a novel approach to network planning and optimization. However, as a technique in its infancy, it admits (and requires) further improvements and evolutions. Hence, we continue the work on this research problem by enriching the possibilities of the original framework (specifically  Steps~1~and~3, Fig.~\ref{Fig:PROP_FRAMEWORK}), and by integrating new tools~(Step~4). To be precise, the contributions of this paper can be summarized as follows: 
\begin{itemize}[leftmargin=0.75cm]
	\renewcommand{\labelitemi}{$\checkmark$} 
	\item The direct mapping (Step~1 in Fig.~\ref{Fig:PROP_FRAMEWORK}) is generalized using function composition~\cite{shapiro2012composition} to overcome the limitations of the mapping proposed in~\cite{7399421}. Thus, the new direct mapping includes both conformal and non-conformal transformations, and allows defining arbitrary spatial service demand distributions in the physical domain. 
	\item A new method for computing the inverse mapping~(Step~3 in Fig.~\ref{Fig:PROP_FRAMEWORK}) using the centroidal Voronoi algorithms~\cite{du1999centroidal} and power Voronoi diagrams~\cite{Tan2011} is presented and explained in this context. As it will be explained later on, this approach does not require the computation of the direct mapping $F$. 
	A comparative analysis between this technique and spatial mappings is also provided.
	\item Power optimization~(Step~4 in Fig.~\ref{Fig:PROP_FRAMEWORK}) to adjust the configuration of the network topologies in the physical domain is integrated into the framework. To the best of our knowledge, the proposed optimization formulation, although based on the existing load-coupling model~\cite{7340975}, is novel and has some convenient features from the planning point~of~view.
\end{itemize}
%The examples presented herein illustrate the use of the proposed enhancements, and the results clearly show their effectiveness of the proposed enhancements, and the effectiveness of the new tools now available for it. Examples and use cases have been selected to avoid any overlapping with the material previously presented~\cite{7399421}, but in such a way that the focus is placed on the novelties, and that they can be presented in a comprehensive manner. 
 
The rest of the paper is organized as follows: the next section provides a high level description of the enhanced framework. Generalizations and novelties for the spatial mappings are presented in Sections~\ref{Sec:DirectMapping}~and~\ref{Sec:InverseMapping}. The analysis in the canonical domain is, for the sake of completeness, briefly described in Section~\ref{Sec:AnalysisInCanDom}; it can be done following the methodology presented in~\mbox{\cite[\textsection$\,$IV]{7399421}.} The proposed power optimization is described in Section~\ref{Sec:PO}. Numerical examples are presented in Section~\ref{Sec:NR}. Section~\ref{Sec:Conclusions} closes the paper with conclusions and future research directions.

 		\section{Framework Description and System Model}\label{Sec:PPSM}
	In general terms, planning aims at determining the number and location of base stations and their corresponding cell areas. The research framework under consideration, shown in Fig.~\ref{Fig:PROP_FRAMEWORK}, is motivated by the notion of service demand and capacity provision compatibility~\cite{04:00389}, in which more access points (with smaller cell areas) are required where the service demand is concentrated. Thus, if base stations have the same amount of resources, the ideal system should be planned and configured so that cells are equally loaded. However, determining such network topology and configuration is not an easy task because, in practice, the service demand is non-uniformly distributed in the coverage area.

		In the proposed framework, two domains are considered. A \textsl{physical} domain~$\mathcal{A}$ that corresponds to the real-world, and a dual \textsl{canonical} domain represented by a rectangular area~$\mathcal{R}$. The service demand in the physical domain is assumed to be known in statistical terms, i.e., its spatial distribution given by a probability density function~$\delta$ defined over $\mathcal{A}$, such that $\int_{\mathcal{A}}\delta(a)\,da=1$, and a certain volume~$V$, expressed in terms of the average number of users, are known. Evidently, both $\delta$ and $V$ vary over time, but for planning purposes, it can be fairly assumed that a given $\delta$ and $V$ (well-known by operators) capture the traffic behavior in representative periods of time~\cite{6381046, 6757900}, i.e., morning, peak-hour, afternoon, and so on.
		
As in~\cite{7399421}, the target is to find a network topology~$\mathcal{T}$ (site's locations and cell areas) with a certain configuration (e.g., power allocation) that is compatible with the service demand and satisfies both coverage and capacity requirements. The proposed framework is composed of the four main steps indicated in Fig.~\ref{Fig:PROP_FRAMEWORK} and explained next:
\begin{enumerate}[leftmargin=*]
	\item \textbf{Direct mapping}. The objective is to determine a mapping function $F:\mathcal{A}\rightarrow\mathcal{R}$, such that the service demand that is non-uniformly distributed (according to $\delta$) in $\mathcal{A}$ becomes uniform in $\mathcal{R}$. The uniform service demand distribution in $\mathcal{R}$ is denoted by $\delta_{\rm u}$, and hence, $\delta_{\rm u}(r)=\frac{1}{|\mathcal{R}|},~\forall\,r\in\mathcal{R}$. Therefore, the function $F$ must be a function of $\delta$. The enhanced direct mapping (Step 1) is discussed in Section~\ref{Sec:DirectMapping}.
	\item \textbf{Analysis in the canonical domain}. The goal is to determine the number~$L$ of uniformly distributed access points required to satisfy the uniformly distributed service demand in~$\mathcal{R}$. Hence, $L$ depends on $V$, i.e., the larger the volume, the higher the density of the uniform topology $\mathcal{T}_{\rm u}$. Note that the same volume $V$ is considered in both domains. A description of the analysis in the canonical domain (Step~2) is provided in Section~\ref{Sec:AnalysisInCanDom}.
	\item \textbf{Inverse mapping}. The target is to find a spatial transformation $F^{-1}:\mathcal{R}\rightarrow\mathcal{A}$ to map the access points from the canonical domain onto the physical domain. If $F$ has been defined, $F^{-1}$ corresponds to its inverse function, and hence, it also depends on $\delta$. However, the mapping~$F^{-1}$ can also be obtained without $F$, directly from $\delta$, as it is illustrated later on. In any case, the inverse mapping conveys the information \textit{stored} in $\delta$ (the spatial service demand distribution). On the one hand, the mapping~$F^{-1}$ guarantees that the resulting network topology~$\mathcal{T}$ in the physical domain is spatially compatible with the service demand, and on the other hand, the~$L$ access points, previously calculated for $\mathcal{T}_{\rm u}$, provide the required capacity to deal with the service demand volume. The options for the computation of the inverse mapping (Step~3) are presented in Section~\ref{Sec:InverseMapping}.
	\item \textbf{Power optimization}. Depending on the non-uniformity of $\delta$ (how much traffic is concentrated in the areas with high demand) and the network density (proportional to $L$), some power optimization is required to \textit{compensate} the interference in \textit{hot-spots}. As indicated, regions with higher demand require smaller cells, and hence, the transmit power needs to be adjusted to avoid unfeasible interference levels, and to equalize the load of the different cells. The proposed power optimization (Step~4) is presented in Section~\ref{Sec:PO}.
\end{enumerate}
 In order to facilitate the reading of the rest of the article, the notation is presented in Table~\ref{Table:Not}.

							\begin{table*}[t]													
		\caption{Notations.}\vspace{-0.5cm}
		\begin{center}
		\begin{tabular}{r l  r l }		
		\toprule  	
			 {\small \textbf{Symbol} }	& {\small\textbf{Description}}	& {\small \textbf{Symbol} }	& {\small\textbf{Description}} 			\\ 
		\midrule 			
						{\small $\mathcal{A}$ }	& {\small Physical domain}	& {\small $\mathcal{R}$} 	& {\small Canonical domain}			\\[-0.04cm]
												{\small $\mathcal{T}_{\rm p}$ }	& {\small Topology in the physical domain}	& {\small $\mathcal{T}_{\rm c}$} 	& {\small Topology in the canonical domain}			\\[-0.04cm]
												{\small $\delta$ }	& {\small Spatial service demand distribution}	& {\small $\delta_{\rm u}$} 	& {\small Uniform service demand distribution}			\\[-0.04cm]
						{\small $L$} 	& {\small Number of base stations}			& {\small $\mathbf{p}\in\mathbb{R}^{L}_{+}$} 	& {\small Power allocation (in data channels)}		\\[-0.04cm]
		{\small $V$} 	& {\small Service demand volume}			& {\small $V_l$} 	& {\small Service demand in the $l^{\rm th}$ cell}		\\[-0.04cm]
				{\small $F$} 	& {\small Direct mapping}			& {\small $F^{-1}$} 	& {\small Inverse mapping}		\\[-0.04cm]
									{\small $\mathbf{p}^{\star}\in\mathbb{R}^{L}_{+}$} 	& {\small Optimized power allocation}			& {\small $B$} 	& {\small System bandwidth}		\\[-0.04cm]
									{\small $R_{\rm min}$} 	& {\small Minimum user rate}			& {\small $\beta$} 	& {\small Propagation exponent}		\\[-0.04cm]
									{\small $\bar{\alpha}_{\rm c}$} 	& {\small Target cell load in $\mathcal{R}$}			& {\small $\bar{\alpha}$} 	& {\small Resulting cell load in $\mathcal{R}^{\prime}$}		\\						
																\bottomrule						 
		\end{tabular} \end{center} 
		\label{Table:Not} 
		\end{table*}		
						
 		\section{Direct Mapping}\label{Sec:DirectMapping}
	\begin{figure*}[t]
	    		\centering	  					
						\subfloat[Space deformation produced by the mapping.]
	    		{\label{Fig:DirectMappingA}
	    		\includegraphics[width = 0.45\textwidth]{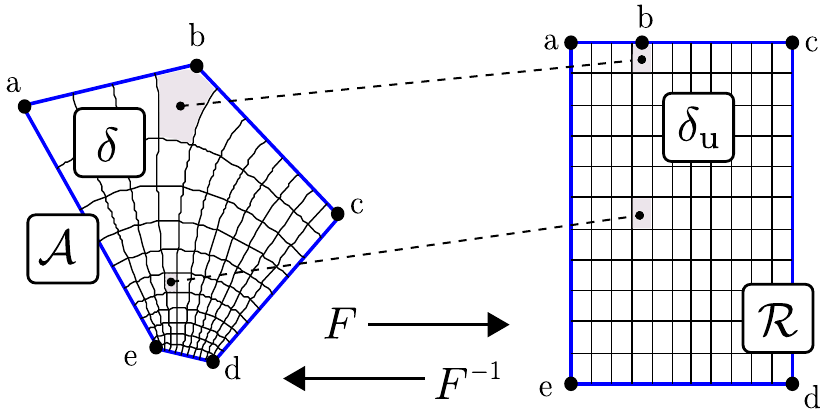}}	\hspace{0.15cm}		
						\subfloat[Mapping composition: polygon $\leftrightarrow$ rectangle.]
	    		{\label{Fig:DirectMappingB}
	    		\includegraphics[width = 0.45\textwidth]{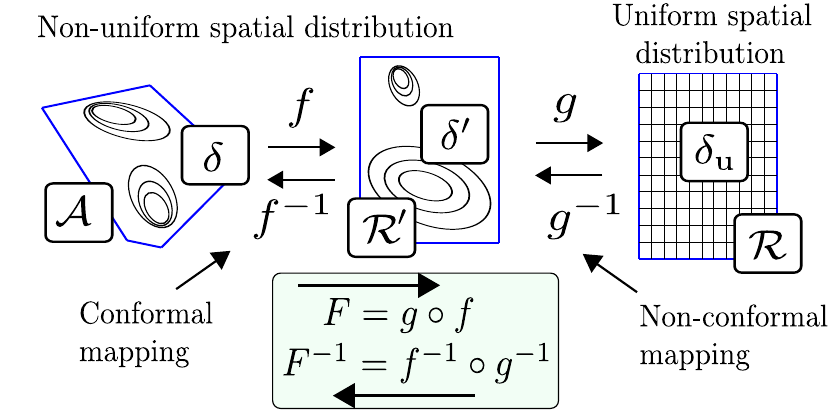}}	
	    		\vspace{-0.3cm}\caption{The direct mapping re-distributes the service demand uniformly in the canonical domain.}
	    		\label{Fig:DirectMapping}    		
\end{figure*}	
The objective of the direct mapping $F$ is to uniformly distribute the service demand from the physical domain onto the canonical domain. To do that, the mapping is required to \textit{stretch} or \textit{compress} the space according to $\delta$, i.e., if $\delta$ admits high values in some region of $\mathcal{A}$, then that region needs to be mapped onto a larger image in $\mathcal{R}$, and viceversa.  This idea is illustrated in Fig.~\ref{Fig:DirectMappingA}, where the spatial service demand distribution is represented by a grid in both domains. Each subregion of the grid has the same amount of traffic. Note that in $\mathcal{R}$, the grid is uniform, and hence, service demand $\delta_{\rm u}$ is also uniform, while in $\mathcal{A}$ is not, i.e., the traffic is concentrated where the grid is denser. In the example, two regions of each domain are connected through the mappings ($F$ and $F^{-1}$), therefore, they are images of each other. It can be seen that, the regions with high traffic density in $\mathcal{A}$ are mapped to larger regions in $\mathcal{R}$, and viceversa. Thus, the mappings not only connect (bijectively) both domains preserving notions such as locality and proximity, but also carries information about the service demand distribution. In~\cite{7399421}, the analysis through canonical domains was presented by implementing the direct mapping $F$ by means of conformal mapping~\cite{olver2015complex}, to be precise, a mapping from polygon onto rectangle using modified Schwarz-Christoffel transformations~\cite{doi:10.1137/0911054}. In order to proof the concept, $\delta$ was assumed to be proportional to the \textit{natural} deformation of the space produced by the conformal mapping used to connect $\mathcal{A}$ and $\mathcal{R}$. However, despite a certain \textit{flexibility} to model $\delta$ when defining the mapping (by conveniently modifying the boundaries and vertices in $\mathcal{A}$, and selecting the corners of the generalized cuadrilateral~\cite{papamichael2010numerical}), the method has limitations. Thus, an improved solution is provided herein to address the more general case when an arbitrary $\delta$ is defined in the physical domain. Given that, to the best of the knowledge of the authors, there are not known methods to conformally map polygons onto rectangles and at the same time redistribute uniformly an arbitrary spatial service demand distribution defined on it; the solution presented herein essentially divides the problem into two pieces as shown in Fig.~\ref{Fig:DirectMappingB}. The idea is to construct the mapping $F:\mathcal{A}\rightarrow\mathcal{R}$ as a composition of two mappings: an initial conformal mapping $f:\mathcal{A}\rightarrow\mathcal{R}^{\prime}$, and a second one $g:\mathcal{R}^{\prime}\rightarrow\mathcal{R}$, such that $F=g \circ f$. The first mapping focuses on the problem of mapping the physical domain (a given polygon) onto the canonical domain (a rectangle), for which complex analysis, i.e., conformal mapping, is required. In general, an arbitrary spatial service demand distribution~$\delta$ defined over $\mathcal{A}$ will be transformed into another non-uniform distribution $\delta^{\prime}$ defined over $\mathcal{R}^{\prime}$. Therefore, a second mapping $g$ is required to homogenize $\delta^{\prime}$. However, this second mapping has a fundamental difference with respect to the first one: it is a mapping between rectangular domains (indeed, $\mathcal{R}$ and $\mathcal{R}^{\prime}$ can be identical), and hence, complex analysis is not longer required.

The mapping $f$ is identical to the one described in~\mbox{\cite[\textsection$\,$III.B]{7399421}}, that was originally proposed in~\cite{doi:10.1137/0911054}. Details on the computation of $f$ and its inverse $f^{-1}$ can be found therein. Additional useful information on conformal mapping, Schwarz-Christoffel transformations, and equivalence among quadrilaterals can be found in~\cite{papamichael2010numerical, driscoll2002schwarz}.

Focusing on the non-conformal mapping ($g$ and $g^{-1}$, see Fig.~\ref{Fig:DirectMappingB}), it is important to indicate that the main reason for calculating the direct mapping $g$ is to obtain its inverse $g^{-1}$. Recall that in the proposed framework, the inverse mapping is the one used to create the network topology~$\mathcal{T}_{\rm p}$ in the physical domain~(sites positioning) once the analysis in the canonical domain is completed. However, if $g^{-1}$ can be obtained directly by means of $\delta^{\prime}$, then $g$ can also be computed by finding the inverse of $g^{-1}$, but in this case it is not strictly necessary. This is the approach used herein, and hence, the computation of $g^{-1}$ is presented in~Section~\ref{Sec:InverseMapping}.

It should be noticed that since the network topology created in $\mathcal{R}^{\prime}$ and the spatial distribution $\delta^{\prime}$ are both non-uniform, $\mathcal{R}^{\prime}$ can be regarded as an intermediate physical domain in which the use of the power optimization proposed in Section~\ref{Sec:PO} can be illustrated without loss of generality. 
				
 		\section{Analysis in the Canonical Domain}\label{Sec:AnalysisInCanDom}
The analysis in the canonical domain addresses the dimensioning part of the planning problem, i.e., to determine how many access points $L$ are required to cope with the service demand, and more precisely with the given volume $V$. 
A full description of the method can be found in~\mbox{\cite[\textsection$\,$IV]{7399421}}. Nevertheless, a brief description is also provided herein for the sake of completeness. 

One of the key objectives in~\cite{7399421} is to perform and simplify part of the planning process by carrying out part of the \textit{job} in a dual domain, referred to as canonical domain, that is~1)~perfectly regular in terms of the geometry of the network topology,~2) uniform in terms of spatial service demand distribution, and~3)~homogeneous in terms of interference. In this manner, independently of the density (proportional to $L$) that is considered, cells are identical in terms of coverage area, service demand, and received interference, and hence,  the analysis of one single cell suffices. In order to do that, a rectangular domain was selected because it is topologically equivalent to the \textit{flat} torus, a 2D manifold~\cite{sullivan2011conformal}, in which the condition of periodicity can be applied to avoid border effects. Thus, cells become equally loaded and the dimensioning is \textit{conservative} since the \textit{wrap-around} produced by the periodicity implies the worst case in terms of interference. The analysis presented in~\cite{7399421} takes into account the cell load coupling model originally presented in~\cite{6204009}, but simpler assumptions such as full load can also be considered. In summary, the objective is to obtain the number of cells $L$ for the regular network topology ($\mathcal{T}_{\rm c}$ in Fig.~\ref{Fig:PROP_FRAMEWORK}) that is required to provide the required capacity, i.e., to find $L$ given $V$. Evidently, there are several factors/assumptions affecting this relationship, such as type of tessellation (squares, rectangles, or hexagons), minimum user rate $R_{\rm min}$, propagation exponent $\beta$, target cell load $\bar{\alpha}_{\rm c}$, and system bandwidth $B$. Since this part of the analysis is inherited as it is from~\mbox{\cite[\textsection$\,$IV]{7399421}}, interested readers are referred therein for the complete description of this dimensioning task. 	
				
		\section{Inverse Mapping}\label{Sec:InverseMapping}
 In this work, two alternatives for finding $F^{-1}$ are presented. The first option is 
 to express $F^{-1}$ in terms of a composition of two functions as indicated in Fig.~\ref{Fig:DirectMappingB}, i.e., $F^{-1}=f^{-1}\circ g^{-1}$. As mentioned before, the computation of the conformal mapping $f^{-1}$ is described in \cite{doi:10.1137/0911054, 7399421}, and the references therein. Hence, with this approach, the task is reduced to obtain $g^{-1}$. The second alternative is to use an algorithmic solution to estimate $F^{-1}$. In this work, this approach is based on Centroidal Voronoi Tessellations~\cite{du1999centroidal}. The idea and method was originally proposed for planning purposes in~\cite{7391090}\footnote{The \textsl{Short Paper} format used in~\cite{7391090} only allowed to introduce the main idea and basic/preliminary results.}. Both methods are described next. 
	\begin{figure*}[t]
	    		\centering	  					
						\subfloat[Notation and baseline assumptions.]
	    		{\label{Fig:NonConfMappA}
	    		\includegraphics[width = 0.47\textwidth]{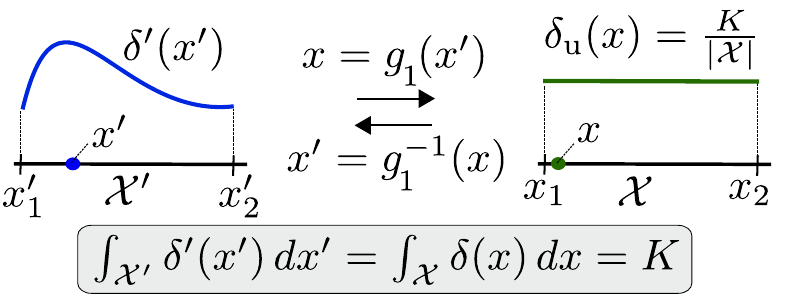}}	\hspace{0.2cm}		
						\subfloat[Mapping in 1D with $\delta^{\prime}(x^{\prime})$ as a linear function.]
	    		{\label{Fig:NonConfMappB}
	    		\includegraphics[width = 0.47\textwidth]{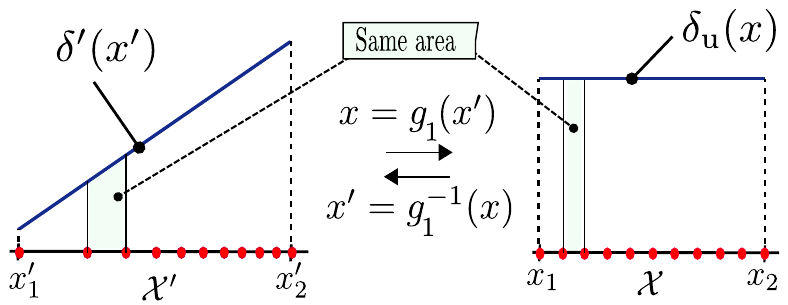}}	
	    		\vspace{-0.3cm}\caption{Required service demand redistribution (mapping) in one dimension.}
	    		\label{Fig:NonConfMapp}    		
\end{figure*}	

\subsection{Non-conformal mapping}\label{Subsec:NonConfMapping}
In order to introduce the proposed mapping, let us to consider the mapping in one dimension $g_1:\mathcal{X}^{\prime}\rightarrow\mathcal{X}$ between two linear domains $\mathcal{X}$ and $\mathcal{X}^{\prime}$ as illustrated in Fig.~\ref{Fig:NonConfMappA}. The density function~$\delta^{\prime}\in\mathbb{R}_{+}$ is defined over $\mathcal{X}^{\prime}$ and the mapping $g_1$ is such that it uniformly re-distributes that \textsl{demand} in $\mathcal{X}$. Its inverse, $g_1^{-1}:\mathcal{X}\rightarrow\mathcal{X}^{\prime}$, on the other hand, must create $\delta^{\prime}$ when mapping a uniform density $\delta_{\rm u}$ from $\mathcal{X}$ onto $\mathcal{X}^{\prime}$. Note that, if $K=1$, $\delta^{\prime}$ and $\delta_{\rm u}$ in Fig.~\ref{Fig:NonConfMappA} can be regarded as probability density functions. Therefore, without loss of generality, hereafter $K$ is assumed to be equal to~$1$.  The mapping must fulfill the following conditions:~1)~the total volume in each domain must be preserved, and hence, \mbox{$\int_{\mathcal{X}^{\prime}}\delta^{\prime}(x^{\prime})\,dx^{\prime}=\int_{\mathcal{X}}\delta_{\rm u}(x)\,dx=1$}, and~2)~$x_1=g_1(x_1^{\prime})$ and $x_2=g_1(x_2^{\prime})$. Intuitively, the function $g_1^{-1}$ has to map the points from $\mathcal{X}$ onto $\mathcal{X}^{\prime}$ in such a way that the area between two mapped points is the same in both domains as it is shown in Fig.~\ref{Fig:NonConfMappB}, where $\delta^{\prime}$ is assumed to be linear. Thus, the space is \textit{compressed} in $\mathcal{X}^{\prime}$ if $\delta^{\prime}$ is high and \textit{streched} is $\delta^{\prime}$ is low. In Fig.~\ref{Fig:NonConfMappB}, red points are images of each other to illustrate this idea. The required mapping $x^{\prime}=g_1^{-1}(x)$ can be obtained by solving for $x^{\prime}$ in the following expression:
		\begin{equation}
		\int_{x_1^{\prime}}^{x^{\prime}}\delta^{\prime}(x^{\prime})\,dx^{\prime}=\int_{x_1}^{x}\delta_{\rm u}(x)\,dx=\frac{x-x_1}{x_2-x_1}.
		\label{Eq:Map1D_1}
	\end{equation}	 
Depending on $\delta^{\prime}$, $x^{\prime}=g_1^{-1}(x)$ can be expressed in closed form, but in general the mapping $g_1^{-1}$ can be evaluated numerically.

\begin{figure}[t]
	    		\centering	    	
	    		\includegraphics[width = 0.51\textwidth]{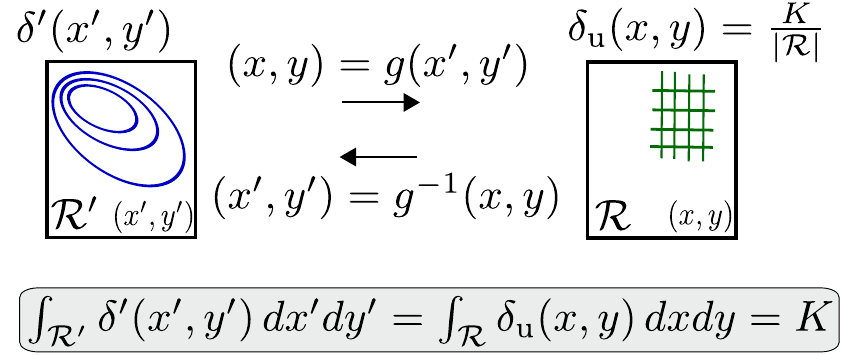}
	    		\vspace{-0.65cm}\caption{Required service demand redistribution (mapping) in two dimensions.}
	    		\label{Fig:NonConfMapp2D}    		
			\end{figure}

			In two dimensions, the goal of the mapping is the same, i.e., to uniformly distribute a non-uniform service demand (volume) from a rectangular domain~$\mathcal{R}^{\prime}$ onto another rectangular domain~$\mathcal{R}$, as shown in the Fig.~\ref{Fig:NonConfMapp2D}. In this case, there are two possibilities: 1)~the spatial service demand distribution~$\delta^{\prime}$ can be expressed as a product of two independent functions of $x^{\prime}$ and $y^{\prime}$ ($\delta^{\prime}_{\rm x}(x^{\prime})$ and $\delta^{\prime}_{\rm y}(y^{\prime})$, respectively), as  follows: $\delta^{\prime}(x^{\prime},y^{\prime})=\delta^{\prime}_{\rm x}(x^{\prime})\,\delta^{\prime}_{\rm y}(y^{\prime})$, i.e., there is statistical independence between $x^{\prime}$ and $y^{\prime}$, and 2)~$\delta^{\prime}(x^{\prime},y^{\prime})$ is given as a joint probability density function that cannot be expressed as a product of independent functions of $x^{\prime}$ and $y^{\prime}$.
In the first case the mapping of each coordinate can be obtained independently by considering each dimension as the one-dimensional (linear) problem previously discussed, i.e., obtaining the mapping through~(\ref{Eq:Map1D_1}). Thus, the mapping $g^{-1}$ can be written as follows:
		\begin{equation}
		(\,x^{\prime},\,y^{\prime}\,) = \left(\,u(x),\,v(y)\,\right),
		\label{Eq:Map2D_SI}
	\end{equation}	
	where $x^{\prime}$ only depends on $x$ and $y^{\prime}$ only depends on $y$. 
	
	If there is no statistical independence between the coordinates, there is no direct way to obtain the required mapping directly from $\delta^{\prime}(x^{\prime},y^{\prime})$. However, the solution for the one-dimensional case, using \eqref{Eq:Map1D_1}, can be employed if the joint probability density function is marginalized~\cite{johnson2011miller}. In this case, the required mapping is not unique as it depends on the variable that is selected first, but in any case it fulfills the requirement of preserving the service demand volume between regions of both domains that are images of each other. The required mapping (assuming without loss of generality that the mapping for $x^{\prime}$ is taken first) can be obtained as follows:	
	\begin{enumerate}
		\item Marginalize $\delta^{\prime}(x^{\prime},y^{\prime})$ with respect to $y^{\prime}$ to obtain the function $\delta^{\prime}_{\rm x}(x^{\prime})$, which is given by
				\begin{equation}
		\delta^{\prime}_{\rm x}(x^{\prime})=\int\delta^{\prime}(x^{\prime},y^{\prime})\,dy^{\prime}. \label{Eq:NonIndeMap_1}
		\end{equation}
		\item Obtain the mapping $x^{\prime}=u(x)$ by means of \eqref{Eq:Map1D_1} and $\delta^{\prime}_{\rm x}(x^{\prime})$.
		\item Use $\delta^{\prime}_{\rm x}(x^{\prime})$ to build $\delta^{\prime}_{\rm y}(y^{\prime},x)$ as follows:
		\begin{equation}
		\delta^{\prime}_{\rm y}(y^{\prime},x)=\frac{\delta^{\prime}(x^{\prime},y^{\prime})}{\delta^{\prime}_{\text{x}}(x^{\prime})} = \frac{\delta^{\prime}(u(x),y^{\prime})}{\delta^{\prime}_{\text{x}}(u(x))}. \label{Eq:NonIndeMap_2}
		\end{equation}
		\item Obtain the mapping $y^{\prime}=v(x,y)$ by means of \eqref{Eq:Map1D_1} and $\delta^{\prime}_{\rm y}(y^{\prime},x)$.
	\end{enumerate}
Thus, the mapping $g^{-1}$ can be written as follows:
\begin{equation}
		(\,x^{\prime},\,y^{\prime}\,)=\left(\,u(x),\,v(x,y)\,\right).\label{Eq:NonIndeMap1_M}
		\end{equation}
Analogously, and following the previous procedure, if $y^{\prime}$ is taken first, the mapping $g^{-1}$ would be given by
\begin{equation}
		(\,x^{\prime},\,y^{\prime}\,)=\left(\,u(x,y),\,v(y)\,\right).\label{Eq:NonIndeMap2_M}
		\end{equation}		
	In both cases, in contrast to (\ref{Eq:Map2D_SI}), one of mappings (either $u$ or $v$) is a function of two variables as indicated in (\ref{Eq:NonIndeMap1_M}) and (\ref{Eq:NonIndeMap2_M}).	
\subsection{A Method based on Centroidal Voronoi Tessellations}\label{Subsec:CVT}
	A Voronoi diagram is a partition of a domain $\mathcal{A}\subset\mathbb{R}^2$ into~$L$ regions ($\mathcal{A}_l,\,l=1,2,\cdots,L$) that are associated to a subset of distinct points $\mathcal{L}=\{\boldsymbol{a}_1,\,\boldsymbol{a}_2,\,\cdots,\,\boldsymbol{a}_L\}\subset\mathcal{A}\subset\mathbb{R}^2$ where 
	\begin{equation}
		\mathcal{A}_l\triangleq\{\boldsymbol{a}\in\mathcal{A}\,|\,\|\boldsymbol{a}-\boldsymbol{a}_l\|_2\leq\|\boldsymbol{a}-\boldsymbol{a}_k\|_2,\forall\,l\neq k\}.\label{Eq:VoronoiRegion}
		\end{equation}
	The notation $\|\,\boldsymbol{a}_1-\boldsymbol{a}_2\,\|_2$ (the $L_2$-norm) indicates the Euclidean distance between the points $\boldsymbol{a}_1$ and $\boldsymbol{a}_2$. The points in $\mathcal{L}$ are known as \textit{generators} of the Voronoi diagram.  
	Voronoi diagrams have extensively been used in the analysis of cellular networks~\mbox{\cite{6399149, 7037539, 6708046}} because the definition of the Voronoi regions is consistent with the coverage areas (cells) of different Base Stations~(BSs), i.e., under the assumption that BSs transmit \textit{pilots} (used for cell selection~\cite{08:00004}) with the same power, cells would correspond to regions defined by~(\ref{Eq:VoronoiRegion}).
If pilots are transmitted with different power levels, power Voronoi diagrams~\cite{Tan2011} can be used to determine cell regions. Power Voronoi diagrams can be regarded as a generalization of classic Voronoi diagrams by assigning \textit{weights} to the generators (in this case BSs), thus locally defining the distance metric. The relation between the pilots' transmit power can be captured by the weights. In Power Voronoi diagrams, cell regions are defined as follows~\cite{Tan2011}: 
		\begin{equation}
		\mathcal{A}_l\triangleq\{\boldsymbol{a}\in\mathcal{A}\,|\,\|\boldsymbol{a}-\boldsymbol{a}_l\|_2-w_l\leq\|\boldsymbol{a}-\boldsymbol{a}_k\|_2-w_k,\forall\,l\neq k\}.\label{Eq:PowerVoronoiRegion}
		\end{equation}
		A \textsl{network topology} ($\mathcal{T}$) in the domain $\mathcal{A}$ is defined as a set of BSs whose locations are indicated by the points \mbox{$\mathcal{L}=\{\boldsymbol{a}_1,\boldsymbol{a}_2,\,\cdots,\,\boldsymbol{a}_L\}\subset\mathcal{A}$} and their corresponding cells ($\mathcal{A}_l$'s) are obtained by means of~(\ref{Eq:PowerVoronoiRegion}), with the weights \mbox{$\mathcal{W}=\{w_1,w_2,\,\cdots,\,w_L\}\subset\mathbb{R}$}. Thus, $\mathcal{T}\triangleq\{\,\mathcal{L},\,\,\mathcal{A}_{l=1,\cdots,L}\,\}$.
		
		The idea introduced in~\cite{7391090} is the joint use of Centroidal Voronoi Tessellations~\cite{du1999centroidal} and Power Voronoi diagrams~\cite{Tan2011} to obtain the topology $\mathcal{T}_{\rm p}$ (see Fig.~\ref{Fig:PROP_FRAMEWORK}) from the number of cells $L$ (obtained through the analysis in the canonical domain, Section~\ref{Sec:AnalysisInCanDom}) and $\delta$; thus achieving the goal of the mapping $F^{-1}$, i.e., to place the $L$ access points in $\mathcal{A}$ in a compatible manner with the spatial service demand distribution.
		
		Before introducing the proposed heuristic, two notions are required: \textsl{tessellation} and \textsl{mass centroid}. A tessellation defined in a domain $\mathcal{A}$ corresponds to a set of regions \mbox{$\mathcal{A}_l\subset\mathcal{A},\,l=1,\cdots,L$,} such that $\cup\mathcal{A}_l=\mathcal{A}$ and $\cap\mathcal{A}_l=\emptyset$. Thus, the sets defined by (\ref{Eq:VoronoiRegion}) and (\ref{Eq:PowerVoronoiRegion}) correspond to tessellations in $\mathcal{A}$. The mass centroid $\boldsymbol{c}_l$ of a region $\mathcal{A}_l\subset\mathcal{A}$ (evidently, $\boldsymbol{c}_l\in\mathcal{A}_l$) is defined as follows:
			\begin{equation}
		\boldsymbol{c}_l\triangleq\frac{\int_{\mathcal{A}_l}\boldsymbol{a}\,\delta(\boldsymbol{a})\,d\boldsymbol{a}}{\int_{\mathcal{A}_l}\delta(\boldsymbol{a})\,d\boldsymbol{a}}=\left(\frac{\iint_{\mathcal{A}_l}x\,\delta(x,y)\,dx\,dy}{\iint_{\mathcal{A}_l}\delta(x,y)\,dx\,dy},\,\frac{\iint_{\mathcal{A}_l}y\,\delta(x,y)\,dx\,dy}{\iint_{\mathcal{A}_l}\delta(x,y)\,dx\,dy}\right),\label{Eq:MassCentroid1}
		\end{equation}	
where $\delta$ is a density defined over $\mathcal{A}$. In this context, $\delta$  corresponds to the spatial service demand distribution (see Fig.~\ref{Fig:PROP_FRAMEWORK}). 

In centroidal Voronoi based algorithms, the idea is to start with an initial set of generators $\mathcal{L}^{\rm R}=\{\boldsymbol{a}_1^{\rm R},\boldsymbol{a}_2^{\rm R},\,\cdots,\,\boldsymbol{a}_L^{\rm R}\}\subset\mathcal{A}$, which can be selected randomly, and compute the corresponding Voronoi diagram and mass centroids $\boldsymbol{c}_l$ of each cell $\mathcal{A}_l$ according to~(\ref{Eq:MassCentroid1}). Then, at each iteration, the~$L$ centroids $\mathcal{C}^{i}=\{\boldsymbol{c}_0^{i},\boldsymbol{c}_1^{i},\cdots,\boldsymbol{c}_L^{i}\}$ of the $i^{\rm th}$ iteration are used as generators of the next Voronoi diagram, i.e., $\mathcal{L}^{i+1}\leftarrow\mathcal{C}^{i}$, till $\boldsymbol{a}_l^{i+1}=\boldsymbol{c}_l^{i},\,\forall l$. The iterative mechanism is required because, in general, Voronoi generators and mass centroids do not match. Hence, the objective is to repeat this process till the generators of the Voronoi diagrams and mass centroids are the same~\cite{du1999centroidal}.

 If the spatial service demand is non-uniform, the centroidal Voronoi algorithm concentrates the access points where the demand is concentrated, i.e., where $\delta$ is high. Thus, the generators of the last Voronoi diagram (and their cells) define a network topology that is compatible with the service demand and that can be used for planning purposes. However, in~\cite{7391090} an improvement based on Power Voronoi diagrams was proposed to obtain network topologies with the service demand uniformly distributed among cells, i.e., $V_l\approx V_k,\,\forall\,l\neq k$. The service demand in the~$l^{\rm th}$~cell is given by
			\begin{equation}
V_l=V\,\int_{\mathcal{A}_l}\delta(\boldsymbol{a})\,d\boldsymbol{a}.\label{Eq:ServDemandVol}
		\end{equation}	
Thus, for a given network topology $\mathcal{T}$, the service demand share $\mathcal{V}$ is defined as follows: $\mathcal{V}=\{\,V_1,V_2,\cdots,V_L\,\}$.
In general, centroidal Voronoi algorithms do not produce network topologies with uniform service demand share; however, power Voronoi diagrams allow independent calibration of cells by adjusting weights, as in (\ref{Eq:PowerVoronoiRegion}). Thus, Algorithm~\ref{Alg:IterSitesLoc} employs both mass centroids and power Voronoi diagrams to obtain network topologies in which the service demand is uniformly distributed among cells, which requires network topologies with different cell sizes. 

Algorithm~\ref{Alg:IterSitesLoc} requires as inputs a random set of $L$ points $\mathcal{L}^{\rm R}\subset\mathcal{A}$, a density function $\delta$, and a set of parameters that controls the operation of the algorithm as described next. Line~\ref{Alg_3} indicates the execution of~$N$ iterations of the \textit{standard} Centroidal Voronoi Algorithm~\cite{du1999centroidal} in order to obtain a \textit{good} starting point for the loop between Lines~\ref{Alg_4}~and~\ref{Alg_11}. At each iteration, the centroids are recalculated from the previous network topology (Line~\ref{Alg_4}), generators are updated~(Line~\ref{Alg_5}), power Voronoi diagrams are computed~(Line~\ref{Alg_6}), and the service demand share $\mathcal{V}$ of the new topology is calculated~(Line~\ref{Alg_7}).  At each iteration, the weight of the cell with the highest demand is updated~(Line~\ref{Alg_9}) and the reduction factor $\Delta$ is updated every $T$ iterations to control the stability/convergence of the algorithm~(Line~\ref{Alg_10}). The algorithm finishes when the Coefficient of Variation~(CoV)\footnote{The CoV is a standardized measure of dispersion and it is defined as the ratio of the standard deviation to the mean.} of the service demand share is smaller than a threshold $\epsilon$~(Line~\ref{Alg_Condition}), i.e., the service demand is well distributed among cells. As it will be seen, Algorithm~\ref{Alg:IterSitesLoc} allows the adjustment of network topologies to homogenize the service demand share; and indeed, a simple variation of it (without updating the site locations) could also be used for fixed/existing networks.
			\begin{algorithm}[t]
 \SetAlgoLined
								\SetKwInOut{Input}{Inputs}
								\SetKwInOut{Output}{Output}	
								\SetKwFunction{CoV}{CoefficientOfVariation}	
								\SetKwFunction{PowerVoronoiDiagram}{PowerVoronoiDiagram}
								\SetKwFunction{Share}{ServiceDemandShare}
								\SetKwFunction{MaxIndex}{MaxIndex}
								\SetKwFunction{mod}{mod}
								\SetKwFunction{MassCentroids}{MassCentroids}
								\SetKwFunction{SCVTA}{CVA}
										\vspace{0.1cm}
							\Input{{\small
										Random network topology: $\mathcal{L}^{\rm R}=\{\boldsymbol{a}_1^{\rm R},\boldsymbol{a}_2^{\rm R},\,\cdots,\,\boldsymbol{a}_L^{\rm R}\}\subset\mathcal{A}$, spatial service demand distribution: $\delta$, algorithm parameters: $N\in\mathbb{N}$, $\Delta<0$, $0<\kappa<1$, $T\in\mathbb{N}$, $\epsilon$.\\
										}} 	\vspace{-0.23cm}								
							\Output{{\small
											Network topology compatible with $\delta$: $\mathcal{T}_{\rm p}$.
										}}
%\BlankLine
\small
$\mathcal{W}\leftarrow\boldsymbol{0}$\tcc*[r]{{\footnotesize Initialization}}\label{Alg_1}\vspace{-0.2cm}

$i\leftarrow1$\label{Alg_2}\;\vspace{-0.2cm}

${\mathcal{T}}^{0}\leftarrow$\SCVTA{ $\mathcal{L}^{\rm R}$, $\delta$, $N\,$ }\tcc*[r]{{\footnotesize Baseline topology: std. centroidal Voronoi algorithm}}\label{Alg_3}\vspace{-0.2cm}

\Repeat{$\,\epsilon\,\leq\,$\CoV{$\mathcal{V}^{i}$}\label{Alg_Condition}}
{
\vspace{-0.2cm}
$\mathcal{C}^{i-1}\leftarrow\,$\MassCentroids{$\,\mathcal{T}^{i-1}$, $\,\delta$ }\tcc*[r]{{\footnotesize Centroids}}\label{Alg_4}\vspace{-0.2cm}

$\mathcal{L}^{i}\leftarrow\,\mathcal{C}^{i-1}$\tcc*[r]{{\footnotesize Update generators}}\label{Alg_5}\vspace{-0.2cm}

	   $\mathcal{T}^{i}\leftarrow\,$\PowerVoronoiDiagram{$\,\mathcal{L}^{i}$,$\,\mathcal{W}$}\tcc*[r]{{\footnotesize Power Voronoi diagram: see (\ref{Eq:PowerVoronoiRegion})}}\label{Alg_6}\vspace{-0.2cm}
		
	   $\mathcal{V}^{i}\leftarrow\,$\Share{$\,\mathcal{T}^{i}$,$\,\delta\,$}\tcc*[r]{{\footnotesize Service demand share: see (\ref{Eq:ServDemandVol})}}\label{Alg_7}\vspace{-0.2cm}		
		
		$j\leftarrow\,$\MaxIndex{$\,\mathcal{V}^{i}\,$}\tcc*[r]{{\footnotesize Index of the cell with the highest demand volume}}\label{Alg_8}	\vspace{-0.2cm}	
		
			$w_{j}\leftarrow\,w_{j}\,+\,\Delta\,$\tcc*[r]{{\footnotesize Reduce the weight of cell $j$}}\label{Alg_9}\vspace{-0.2cm}

			\If{ (~\mod{\,$i,T$}==0~) \label{Alg_If}}
			{\vspace{-0.2cm}
			$\Delta\,\leftarrow \Delta\cdot\kappa$\tcc*[r]{{\footnotesize Convergence: reduce $\Delta$ every $T$ iterations}}\label{Alg_10}\vspace{-0.2cm}
			}\vspace{-0.2cm}
		$i\leftarrow i+1$\label{Alg_11}\;\vspace{-0.2cm}			
}\vspace{-0.2cm}
\textbf{return}~~$\mathcal{T}_{\rm p}\,\leftarrow\mathcal{T}^{i}$\tcc*[r]{{\footnotesize Return network topology}}	
		\caption{Network planning based on centroidal and power Voronoi diagrams.}
		\label{Alg:IterSitesLoc}
		\end{algorithm}

		\section{Power Optimization}\label{Sec:PO}
 The need for power optimization is intuitively justified by the fact that, when the spatial service demand is very irregular, access points are concentrated in high demand  areas (to increase the spatial frequency reuse); leading to very high levels of interference if power is not adjusted proportionally to cell sizes. In the context of Orthogonal Frequency Division Multiple Access~(OFDMA) networks, the effect of interference on the \textit{load} of each cell is accurately described by the \mbox{load-coupling} model introduced in~\cite{6204009}. As in~\cite{7399421}, this model is adopted and briefly described here (for the sake of completeness) as a starting point for the optimization framework that is proposed. The load $\alpha$ of a cell is defined as the fraction of resources that is required, on average, to satisfy the service demand. Thus, following~\cite{7399421, 6204009}, the load $\alpha_l$ in the~$l^{\rm th}$ cell is given by
\begin{equation}
	\alpha_l=\frac{V\,R_{\rm min}\,{\rm log}(2)}{B} \int_{\mathcal{A}_l}\frac{\delta_a}{\log{\left(1+\gamma_a(\boldsymbol{\alpha},\boldsymbol{p})\right)}}da. \label{Eq:baseline0}
\end{equation}
%where $\alpha_l$ is the load of the $l^{\rm th}$ cell, i.e., the fraction of resources that are required, on average, to fulfill users' target rate. 
Here, the factor $\frac{V\,R_{\rm min}\,{\rm log}(2)}{B}$ is a constant, $B$ is the system bandwidth, $R_{\rm min}$ is the target rate, $\mathcal{A}_l$ is the coverage of the $l^{\rm th}$ access point, and $\delta_a$ is the relative service demand in the $a^{\rm th}$ area element~(given by the spatial service demand distribution ($\delta$) under consideration). The network coverage area is $\mathcal{A}=\bigcup\mathcal{A}_l$, with $\mathcal{A}_i\cap\mathcal{A}_j=\emptyset,~\forall i\neq j$. The function 
 $\gamma_a(\boldsymbol{\alpha},\mathbf{p})$ is the SIR in the~$a^{\rm th}$~area element. It is a function of the load in other cells and the power allocation, $\boldsymbol{\alpha}\in\mathbb{R}_{+}^L$ and $\mathbf{p}\in\mathbb{R}_{+}^L$, respectively. The SIR can be expressed as follows:
\begin{equation}
\gamma_a(\boldsymbol{\alpha},\mathbf{p}) = \frac{p_{\hat{l}}\,G_{\hat{l},a}}{\,\sum\limits_{l=1,\,{l\neq \hat{l}}}^{L} p_l\,G_{l,a}\,\alpha_l} ,
\end{equation}
where $\hat{l}$ is the index of the serving access point, $p_{l}$ is the transmit power in the $l^{\rm th}$ access point, and $G_{la}$ is the average channel gain between the $a^{\rm th}$ area element and the $l^{\rm th}$ access point. Note that interference coming from neighbors is scaled by the corresponding load factors ($\alpha_l$'s). Thus, the so-called Non-linear Load Coupling Equations~(NLCE), $\boldsymbol{f}:\mathbb{R}^{L}\rightarrow \mathbb{R}_{+}^L$, can be written as follows~\cite{6204009, 6887352}:
\begin{equation}
	\boldsymbol{\alpha}=\boldsymbol{f}(\boldsymbol{\alpha}\,,\,\mathbf{p};\,\delta,\,\mathbf{G},\,V,\,B,\,R_{\rm min}), \label{Eq:baseline}
\end{equation}
where $\int_{\mathcal{A}}\delta_a\,da=1$ and $\mathbf{G}\in\mathbb{R}^{L\times A}$ contains information on the network geometry, i.e., the average channel gain between each area element and access point, and $a=1,2,\cdots,\,A$. Hereafter, for the sake of clarity, (\ref{Eq:baseline}) is simply written as:
\begin{equation}
	\boldsymbol{\alpha}=\boldsymbol{f}(\boldsymbol{\alpha}\,,\,\mathbf{p}). \label{Eq:baselineS}
\end{equation}
As indicated, mathematical properties of (\ref{Eq:baselineS}) including existence and uniqueness of solutions are presented in~\cite{6204009}, under the assumption that $\mathbf{p}$ is given. The recent papers~\cite{6887352, 7340975, 7024780} include the power vector $\mathbf{p}$ as an optimization variable to achieve several goals, such as  minimization of the sum of the loads~($\sum\alpha_l$) or minimization of the transmit power~(the product $\boldsymbol{\alpha}\cdot\mathbf{p}$). Recent improvements in the required algorithmic has also been introduced in~\cite{7032233, 7332797}. All these excellent contributions have increased our understanding about load-coupling in OFDMA-based cellular networks and provide useful optimization frameworks.  

In~\cite{6887352}, for instance, notions such as \textsl{rate satisfiability} and \textsl{load implementability} are developed. Essentially, while for every power allocation $\mathbf{p}$ there is a corresponding load pattern $\boldsymbol{\alpha}$, the converse is not true. The authors present important results regarding the existence and computation of $\mathbf{p}$ for a given load pattern $\boldsymbol{\alpha}$ (as long as $\boldsymbol{\alpha}$ is implementable). To that end, an Iterative Algorithm for Power~(IAP)~\cite{414651} is also presented, and its convergence is shown. It was also proven that $\boldsymbol{\alpha}=\mathbf{1}$ is optimal from the energy efficiency point of view. However, having cells operating at full capacity/load could be not advisable from a practical point of view. 

In this work, and as a part of the framework presented herein, a different power optimization formulation is presented and studied. In order to be aligned with~\cite{04:00389}, where a notion of \textit{irregularity} is defined in terms of the dispersion of the load vector~$\boldsymbol{\alpha}$, the proposed power optimization aims at minimizing the variance of $\boldsymbol{\alpha}$, i.e., ${\rm Var\{\boldsymbol{\alpha}\}}$, using $\mathbf{p}$ as optimization variable. This simple, yet interesting, approach has the following important and convenient features:
\begin{enumerate}
  \item Solving the proposed power optimization results in a uniform load pattern $\bar{\alpha}\cdot\mathbf{1}$, with $\bar{\alpha}\in\mathbb{R}_{+}$. This means that cells are equally loaded. As remarked in~\cite{7462493}, this is a very important target in planning, where distributing the service demand as evenly as possible is desirable. Indeed, the approach provides a better result because having uniform service demand share does not imply a uniformly loaded network. The proposed optimization does provide the power allocation that is required to achieve the aforementioned important network planning target; and it is done taking into account the load-coupling model (including service demand spatial distribution, interference, and so on). In addition, as it is discussed in~\cite{04:00389}, achieving the previous goal also maximizes the service demand volume the network is able to manage. 
	\item No $\boldsymbol{\alpha}$ needs to be specified beforehand. The optimization converges to a resulting uniform load pattern $\bar{\alpha}\cdot\mathbf{1}$, where $\bar{\alpha}$ does not need to be known in advance. 
	\item The \textit{spare} capacity is maximized, thus providing robustness as the maximum network-wide protection against instantaneous traffic variations is obtained.
\end{enumerate}
The proposed power optimization can be written as follows: 
	\begin{eqnarray}
	\underset{\mathbf{p}}{\operatorname{minimize}}&&~~~{\rm Var}\{ \boldsymbol{\alpha} \},\label{OP2}\\
	{\rm subject~to:}&&\nonumber\\[-0.6cm]
	&&\boldsymbol{\alpha}=\boldsymbol{f}(\boldsymbol{\alpha}\,,\,\mathbf{p}),\nonumber\\
	&&\mathbf{p}\in\mathbb{R}^L_{+}.\nonumber
	\end{eqnarray}
	Thus, by applying $\mathbf{p}^{\star}$, the load of each cell becomes equal to $\bar{\alpha}$. Once $\mathbf{p}^{\star}$ is applied, the network load level, i.e., the value of $\bar{\alpha}$, can be modified by varying the variables $V$, $B$, and $R_{\rm min}$. Analogously, any power allocation $\kappa\cdot\mathbf{p}^{\star}$, with $\kappa\in\mathbb{R}_{+}$, is also a solution of~(\ref{OP2}).	In practice, actual levels must consider coverage criteria, as a minimum received power is required. Problem~\eqref{OP2} can be addressed by means of solvers based on interior-point methods~\cite{potra2000interior} or through heuristics, such as IAP~\cite{6887352, 414651}. Convergence and uniqueness aspects are discussed in Appendix~\ref{App:CU}. It is important to point out that solving (\ref{OP2}) provides the power allocation for the data channels, while keeping the cells (the regions $\mathcal{A}_l$) fixed. This can be regarded as a form of load balancing~\cite{6812287} that, in contrast to existing methods based on cell range adjustments (cell-breathing like schemes), do not transfer traffic from one cell to another, but compensate the load pattern by adjusting interference conditions in the network.

 		\section{Numerical Results}\label{Sec:NR}
In this work, a generalization of the direct mapping $F$ (see Fig.~\ref{Fig:PROP_FRAMEWORK}) in terms of the composition of two functions has been presented. As it is explained in Section~\ref{Sec:DirectMapping}, and shown in Fig.~\ref{Fig:DirectMappingB}, the direct mapping $F$ can be written as follows: $F=g \circ f$, where $f$ is a conformal mapping and~$g$~is a non-conformal (rectangle-onto-rectangle) mapping. Given that computation and use of the conformal mapping $f$ (and its inverse $f^{-1}$) has been fully addressed in~\cite{7399421}, the numerical examples illustrate the computation, use, and performance of the new components of the mapping~$F$, i.e., the mappings $g$ and $g^{-1}$ (see Figs.~\ref{Fig:DirectMappingB}~and~\ref{Fig:NonConfMapp2D}) between the domains~$\mathcal{R}$~and~$\mathcal{R}^{\prime}$, in Sections~\ref{Sec:NR:Mappings}~and~\ref{Sec:NR:SD_Share_LoadPatterns}. Section~\ref{Sec:NR:LoadBalancingPO} presents the results regarding power optimization, and finally, a comparative perspective is provided in Section~\ref{Sec:NR:Comparative}.
\subsection{Spatial mappings}\label{Sec:NR:Mappings}
\begin{figure*}[t]
	    		\centering	  					
						\subfloat[$\delta^{\prime}_{1}(x^{\prime},y^{\prime})=x^{\prime}\,e^{-y^{\prime}}$.]
	    		{\label{Fig:NumExa_1}
	    		\includegraphics[width = 0.31\textwidth]{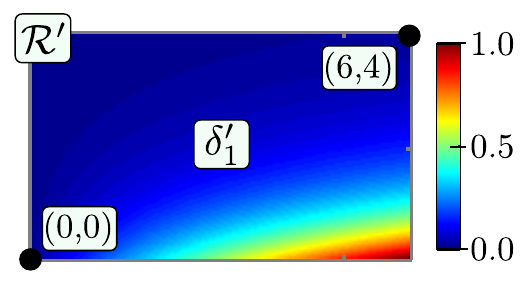}}	\hspace{0.2cm}		
						\subfloat[$\delta^{\prime}_{2}(x^{\prime},y^{\prime})=x^{\prime}+y^{\prime}$.]
	    		{\label{Fig:NumExa_2}
	    		\includegraphics[width = 0.31\textwidth]{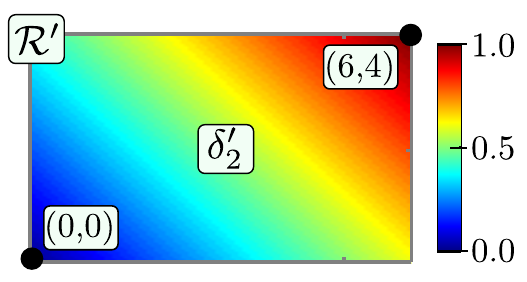}}	
					\subfloat[Synthetic network topology: $\mathcal{T}_{\rm c}$.]
	    		{\label{Fig:NumExa_3}
	    		\includegraphics[width = 0.31\textwidth]{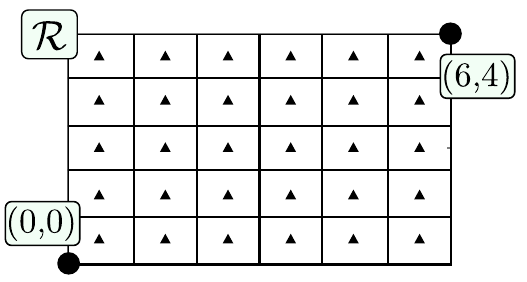}}	\\[-0.2cm]
					\subfloat[Mapping: $\mathcal{T}_{\rm c}\,\overset{\mathclap{g^{-1}_{1{\rm m}}}}{\longrightarrow}\,\mathcal{T}_{\rm 1m}^{\prime}$.]
	    		{\label{Fig:NumExa_4}
	    		\includegraphics[width = 0.24\textwidth]{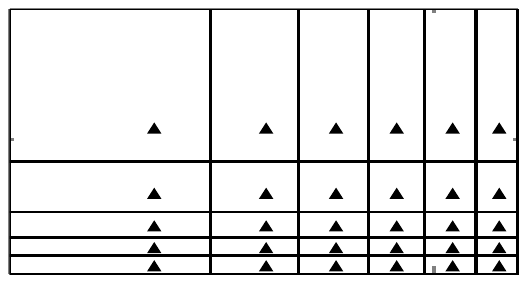}}	
					\subfloat[Mapping: $\mathcal{T}_{\rm c}\,\overset{\mathclap{g^{-1}_{1{\rm c}}}}{\longrightarrow}\,\mathcal{T}_{\rm 1c}^{\prime}$.]
	    		{\label{Fig:NumExa_5}
	    		\includegraphics[width = 0.24\textwidth]{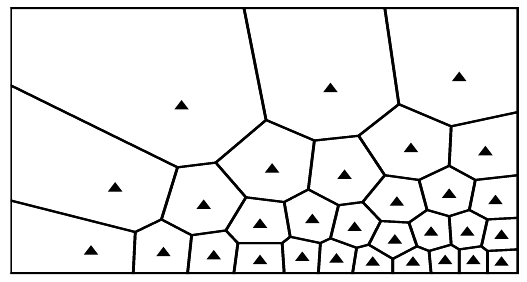}}	
					\subfloat[Mapping: $\mathcal{T}_{\rm c}\,\overset{\mathclap{g^{-1}_{2{\rm m}}}}{\longrightarrow}\,\mathcal{T}_{\rm 2m}^{\prime}$.]
	    		{\label{Fig:NumExa_6}
	    		\includegraphics[width = 0.24\textwidth]{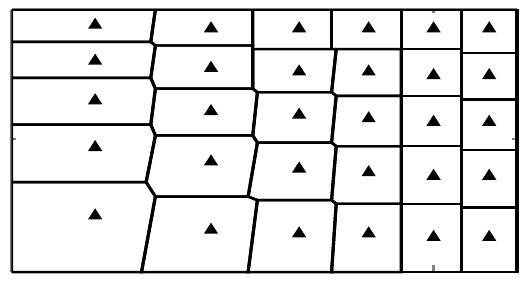}}	
					\subfloat[Mapping: $\mathcal{T}_{\rm c}\,\overset{\mathclap{g^{-1}_{2{\rm c}}}}{\longrightarrow}\,\mathcal{T}_{\rm 2c}^{\prime}$.]
	    		{\label{Fig:NumExa_7}
	    		\includegraphics[width = 0.24\textwidth]{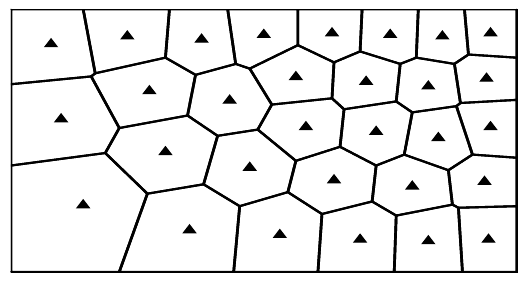}}	
	    		\vspace{-0.3cm}\caption{Examples to illustrate the use of the proposed spatial mappings}
	    		\label{Fig:NumExa}    		
\end{figure*}

% Since the use of conformal mapping was presented in~\cite{7399421}, numerical examples show the calculation of the inverse mapping $g^{-1}$~(see Figs.~\ref{Fig:DirectMappingB}~and~\ref{Fig:NonConfMapp2D}) by means of the methods described in Section~\ref{Sec:InverseMapping}. 
The setting used in the examples is shown in Fig.~\ref{Fig:NumExa}, where Figs.~\ref{Fig:NumExa_1}-\ref{Fig:NumExa_3} illustrate the domains $\mathcal{R}^{\prime}$ and $\mathcal{R}$. The spatial service demand distributions~$\delta^{\prime}_{1}$ and $\delta^{\prime}_{2}$ are represented in Figs.~\ref{Fig:NumExa_1}~and~\ref{Fig:NumExa_2}, respectively; while the synthetic network topology~$\mathcal{T}_{\rm c}$ is shown in Fig.~\ref{Fig:NumExa_3}. The topology $\mathcal{T}_{\rm c}$ was obtained through the dimensioning analysis in the canonical domain (Section~\ref{Sec:AnalysisInCanDom}). For that, the following assumptions were made: 
\begin{itemize}
	\item Service demand volume: $V=90\,{\rm s}/130\,{\rm ms}=692.3$ users (on average); based on the average session time and average inter-arrival time, $90\,{\rm s}$ and $130\,{\rm ms}$, respectively. 
	\item System bandwidth: $B=20.0\,{\rm MHz}$.
	\item Minimum user rate: $R_{\rm min}=1.0\,{\rm Mbps}$.
	\item Average channel gain: $G=d^{-\beta}$; $d$ is the distance and $\beta=3$ is the propagation exponent.
	\item Uniform power allocation: $\mathbf{p}=\mathbf{1}$, and uniform service demand distribution.
\end{itemize}
A target load was $\bar{\alpha}_{\rm c}\approx 0.9$ for the canonical domain. Then, a number of base stations $L=30$ was found using a rectangular cell layout as shown in Fig.~\ref{Fig:NumExa_3}. Recall that the same parameters are assumed for the domain $\mathcal{R}^{\prime}$, but service demand distribution is obtained according to $\delta^{\prime}_{1}$ and $\delta^{\prime}_{2}$. The different mappings of the topology~$\mathcal{T}_{\rm c}$ onto $\mathcal{R}^{\prime}$ are illustrated in Figs.~\ref{Fig:NumExa_4}-\ref{Fig:NumExa_7}. The subscripts of the mappings~$g^{-1}$ are composed of one number~(1~or~2) to refer to the spatial service demand distribution ($\delta^{\prime}_{1}$ or $\delta^{\prime}_{2}$) that is used to create the mapping, and one letter to indicate the method, i.e., `m' for non-conformal mapping (Section~\ref{Subsec:NonConfMapping}) and `c'~for~centroidal Voronoi algorithms~(Section~\ref{Subsec:CVT}). For instance, $g^{-1}_{1{\rm m}}$ (Fig.~\ref{Fig:NumExa_4}) indicates the non-conformal mapping of~$\mathcal{T}_{\rm c}$ from $\mathcal{R}$ onto $\mathcal{R}^{\prime}$, for $\delta^{\prime}_{1}$. The same subscripts are used to refer to the non-uniform topologies~($\mathcal{T}^{\prime}$'s, Figs.~\ref{Fig:NumExa_4}-\ref{Fig:NumExa_7}) created in each case.

Note that $\delta_1^{\prime}$ (Fig.~\ref{Fig:NumExa_1}) corresponds to a case of statistical independence between $x^{\prime}$ and $y^{\prime}$, while $\delta_2^{\prime}$~(Fig.~\ref{Fig:NumExa_2}) does not. Hence, $g^{-1}_{1{\rm m}}$ can be expressed as (\ref{Eq:Map2D_SI}) and $g^{-1}_{2{\rm m}}$ can be expressed as either (\ref{Eq:NonIndeMap1_M}) or (\ref{Eq:NonIndeMap2_M}). 

The mapping~$g^{-1}_{1{\rm m}}$ is given by
		\begin{equation}
	x^{\prime} = u(x) = \sqrt{6\,x},\label{Eq:Map_g1m_x}
		\end{equation}
and
		\begin{equation}
	y^{\prime} = v(y) = -\log\left( 1-\frac{y}{4}\left( 1-e^{-4} \right) \right).\label{Eq:Map_g1m_y}
		\end{equation}

\begin{figure}[t]
	    		\centering	    	
	    		\includegraphics[width = 0.99\textwidth]{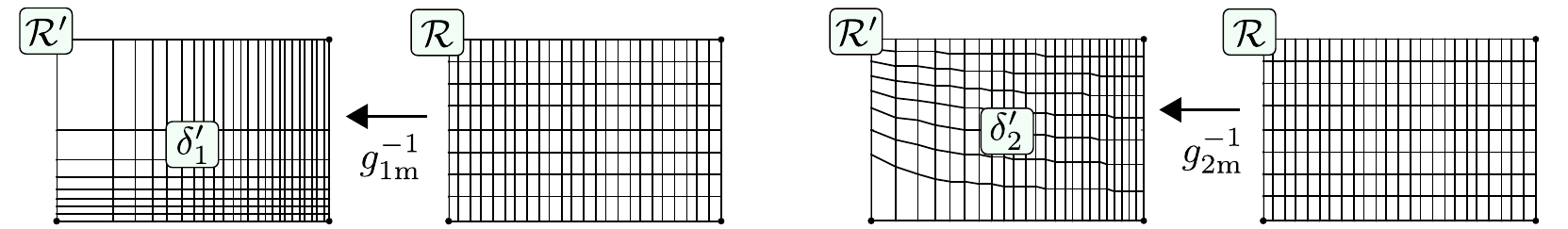}
	    		\vspace{-0.7cm}\caption{Illustration of the non-conformal mappings used in the examples.}
	    		\label{Fig:NonConformalMappings}    		
			\end{figure}

The mapping~$g^{-1}_{2{\rm m}}$ is given by
		\begin{equation}
	x^{\prime} = u(x) = \frac{1}{2}\left( -4 + \sqrt{16 + 40x} \right),\label{Eq:Map_g2m_x}
		\end{equation}
and
		\begin{equation}
	y^{\prime} = v(x,y) = -u(x)+\sqrt{ \left[\,u(x)\,\right]^2 + y\left[\,2\,u(x)+4\,\right] }.\label{Eq:Map_g2m_y}
		\end{equation}
Both $g^{-1}_{1{\rm m}}$ and $g^{-1}_{2{\rm m}}$ are illustrated in Fig.~\ref{Fig:NonConformalMappings}. It becomes clear how the mappings \textit{compress} the space~(in $\mathcal{R}^{\prime}$) where the service demand is high. Thus, we obtain service provision that is compatible with the service demand, i.e., more access points where the demand is concentrated.

\begin{figure*}[t]
	    		\centering	  					
						\subfloat[Topology after standard CVA.]
	    		{\label{Fig:CPV_Oper1}
	    		\includegraphics[width = 0.28\textwidth]{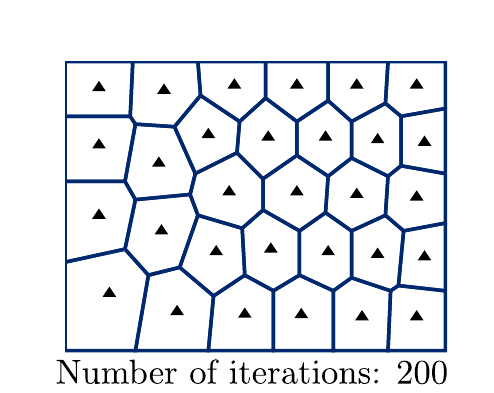}}	\hspace{0.2cm}		
						\subfloat[Topology after Algorithm 1.]
	    		{\label{Fig:CPV_Oper2}
	    		\includegraphics[width = 0.28\textwidth]{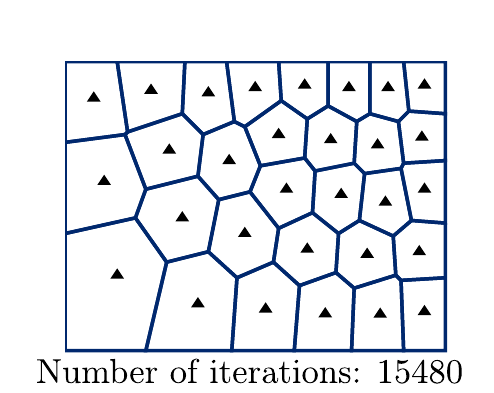}}	
					\subfloat[Share and cells adjustment.]
	    		{\label{Fig:CPV_Oper3}
	    		\includegraphics[width = 0.28\textwidth]{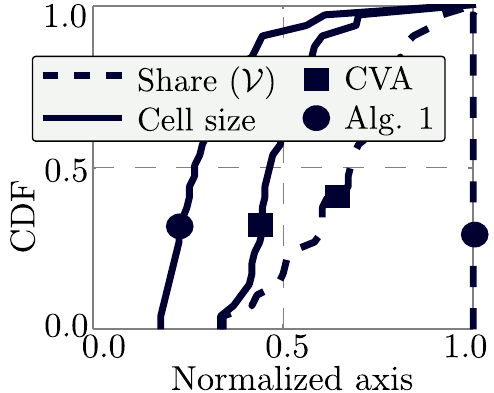}}	\\[-0.025cm]
					\subfloat[Resulting weights.]
	    		{\label{Fig:CPV_Oper6}
	    		\includegraphics[width = 0.28\textwidth]{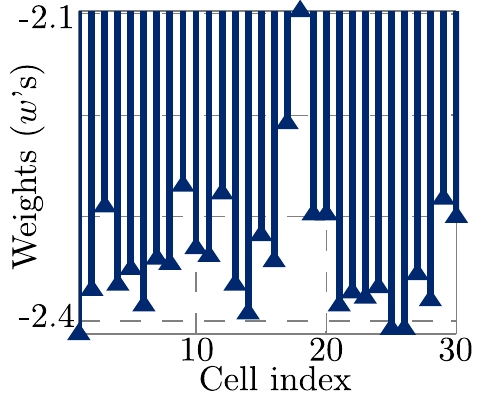}}
					\subfloat[Load balancing.]
	    		{\label{Fig:CPV_Oper4}
	    		\includegraphics[width = 0.28\textwidth]{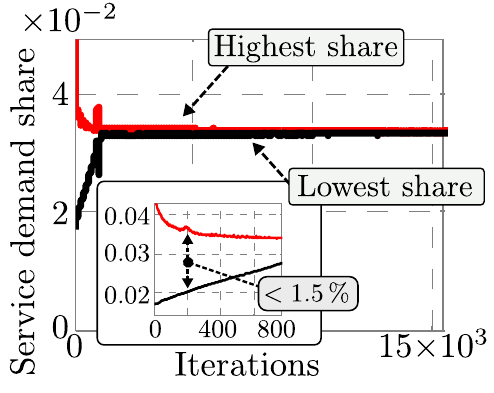}}					
					\subfloat[Reduction of $\Delta$.]
	    		{\label{Fig:CPV_Oper5}
	    		\includegraphics[width = 0.28\textwidth]{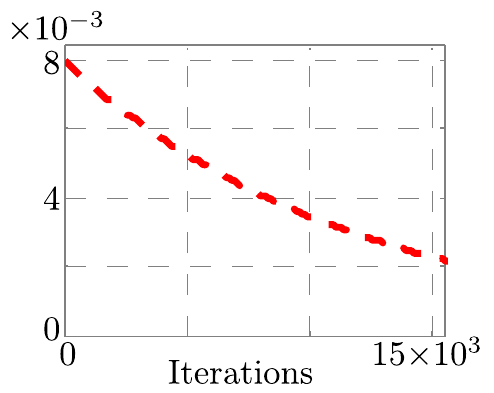}}
	    		\vspace{-0.3cm}\caption{Operation of Algorithm 1 with $\delta_2^{\prime}$.}
	    		\label{Fig:CPV_Oper}    		
\end{figure*}

Regarding the mappings $g^{-1}_{1{\rm c}}$ and $g^{-1}_{2{\rm c}}$, evidently they cannot be expressed in closed form as resulting topologies are obtained after the execution of Algorithm~\ref{Alg:IterSitesLoc}. However, it is recalled that the key step is the calculation of the mass centroids for each cell at each iteration according to~(\ref{Eq:MassCentroid1}). This is the mechanism by which Algorithm~\ref{Alg:IterSitesLoc}
achieves higher density of access points where the service demand is concentrated. The operation of Algorithm~\ref{Alg:IterSitesLoc} is illustrated in Fig.~\ref{Fig:CPV_Oper} using the function $\delta_2^{\prime}$ as example. Similar operation pattern is obtained for~$\delta_1^{\prime}$. 
Figs.~\ref{Fig:CPV_Oper1} corresponds to the network topology obtained after the execution of 200 iterations the standard centroidal Voronoi algorithm (Line~\ref{Alg_3} in Algorithm~\ref{Alg:IterSitesLoc}). In general, centroidal Voronoi algorithms do not result in homogeneous service demand share ($V_l\approx V_k,\,\forall\,l\neq k$), however, they provide a good starting point for Algorithm~\ref{Alg:IterSitesLoc}. The network topology after the execution of Algorithm~\ref{Alg:IterSitesLoc} is shown in Fig.~\ref{Fig:CPV_Oper2}. Fig.~\ref{Fig:CPV_Oper3} shows a comparative perspective between the standard Centroidal Voronoi Algorithm~(CVA) and Algorithm~\ref{Alg:IterSitesLoc}, associated in the figure to squares and circles, respectively. Cumulative Distribution Functions~(CDFs) of cells size (solid patterns) and service demand share (dash patterns) are shown. All the CDFs are normalized, and hence, $x$-axis also goes from~0~to~1. CDFs of cells size indicate diversity in terms of cells area size, and CDFs of service demand share indicate how well distributed among cells the service demand is. Clearly, Algorithm~\ref{Alg:IterSitesLoc} succeeds in achieving uniform service demand share, while CVA does not. As explained earlier, Algorithm~\ref{Alg:IterSitesLoc} 
gradually adjusts the weights of the power Voronoi diagram (Fig.~\ref{Fig:CPV_Oper6}), such that the cell with the highest share reduces its coverage, while cells with less demand tends to increase its coverage. The net result is an increase in cell size range (with respect to CVA), as it is shown in Fig.~\ref{Fig:CPV_Oper3}. Note that the CDF of cells size for the CVA indicates higher cell size homogeneity, which can be verified visually by looking at Figs.~\ref{Fig:CPV_Oper1}~and~\ref{Fig:CPV_Oper2}. Fig.~\ref{Fig:CPV_Oper4} shows the evolution of the highest and lowest cell share in the network. In the example, Algorithm~\ref{Alg:IterSitesLoc} reduces in~200~iterations the gap between the cells with highest and smallest share in less than $1.5\,\%$. The figure illustrates the asymptotic convergence of Algorithm~\ref{Alg:IterSitesLoc}. As mentioned, the reduction factor $\Delta$ (see Algorithm~\ref{Alg:IterSitesLoc}) is gradually decreased to enhance the convergence of the algorithm as shown in Fig.~\ref{Fig:CPV_Oper5}.

%It is important to stress that, in OFDMA-based cellular networks, having uniform service demand share does not imply uniform load mainly because interference is not uniformly distributed and load coupling effects. In practice, less power is allocated to small cells to keep interference under \textit{acceptable} levels. In this work, the optimization formulation provided in Section~\ref{Sec:PO} is used to optimize the power allocation, such that the network becomes uniformly loaded. However, as it will be shown later on, having uniform service demand share is not a sufficient nor necessary condition to obtain cells  uniformly loaded; however, having means to achieve both goals is definitively desirable. 

\subsection{Service demand share and load patterns}\label{Sec:NR:SD_Share_LoadPatterns}
\begin{figure*}[t]
	    		\centering	  					
					\subfloat[Share: $\mathcal{T}_{\rm c}$.]
	    		{\label{Fig:SLP1}
	    		\includegraphics[width = 0.199\textwidth]{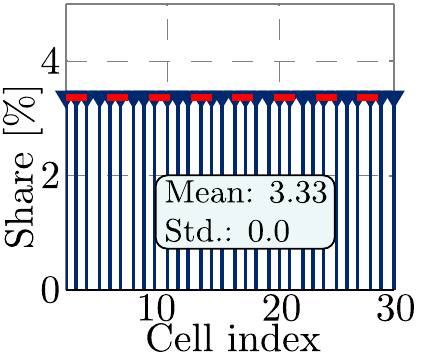}}			
						\subfloat[Share: $\mathcal{T}_{\rm 1m}^{\prime}$.]
	    		{\label{Fig:SLP2}
	    		\includegraphics[width = 0.199\textwidth]{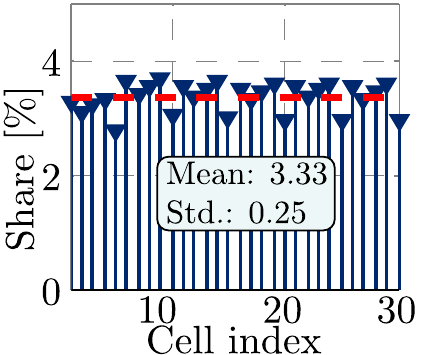}}	
					\subfloat[Share: $\mathcal{T}_{\rm 1c}^{\prime}$.]
	    		{\label{Fig:SLP3}
	    		\includegraphics[width = 0.199\textwidth]{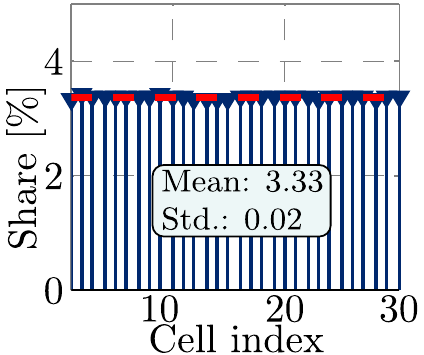}}	
					\subfloat[Share: $\mathcal{T}_{\rm 2m}^{\prime}$.]
	    		{\label{Fig:SLP4}
	    		\includegraphics[width = 0.199\textwidth]{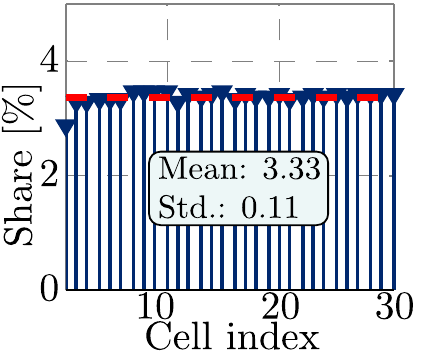}}
					\subfloat[Share: $\mathcal{T}_{\rm 2c}^{\prime}$.]
	    		{\label{Fig:SLP5}
	    		\includegraphics[width = 0.199\textwidth]{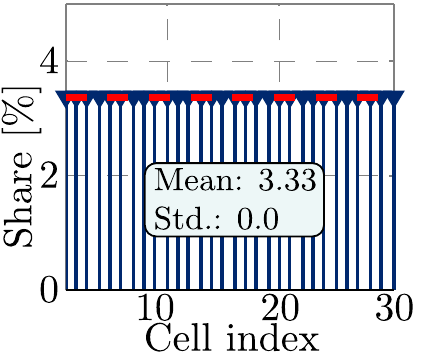}}		\\	
					\subfloat[Load: $\mathcal{T}_{\rm c}$.]
	    		{\label{Fig:SLP6}
	    		\includegraphics[width = 0.199\textwidth]{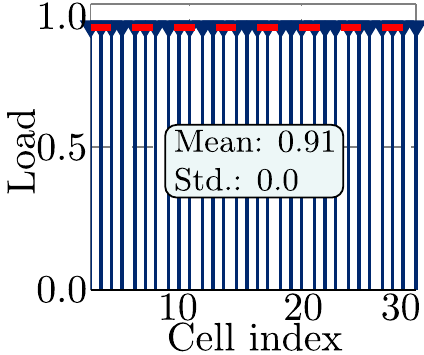}}			
						\subfloat[Load: $\mathcal{T}_{\rm 1m}^{\prime}$.]
	    		{\label{Fig:SLP7}
	    		\includegraphics[width = 0.199\textwidth]{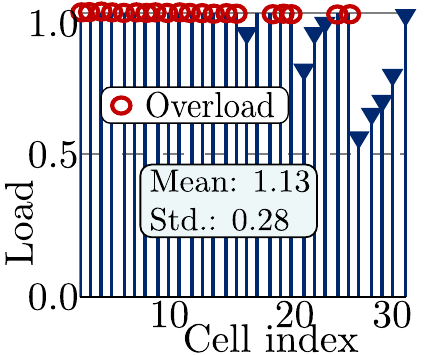}}	
					\subfloat[Load: $\mathcal{T}_{\rm 1c}^{\prime}$.]
	    		{\label{Fig:SLP8}
	    		\includegraphics[width = 0.199\textwidth]{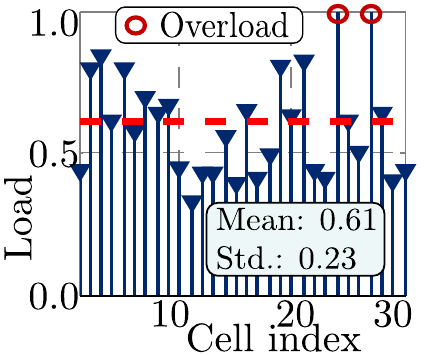}}	
					\subfloat[Load: $\mathcal{T}_{\rm 2m}^{\prime}$.]
	    		{\label{Fig:SLP9}
	    		\includegraphics[width = 0.199\textwidth]{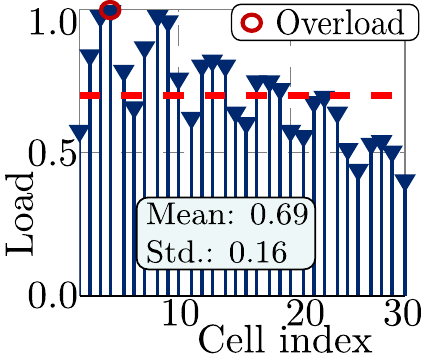}}
					\subfloat[Load: $\mathcal{T}_{\rm 2c}^{\prime}$.]
	    		{\label{Fig:SLP10}
	    		\includegraphics[width = 0.199\textwidth]{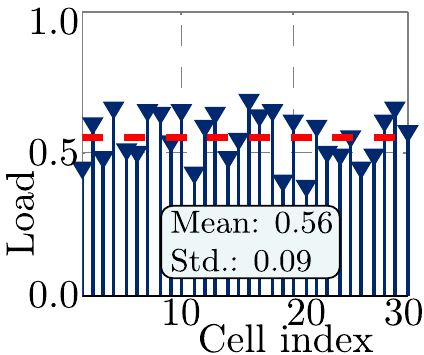}}	
	    		\vspace{-0.3cm}\caption{Service demand share and load pattern (under uniform power allocation).}
	    		\label{Fig:SLP}    		
\end{figure*}

The service demand share and load patterns of the topologies illustrated in Figs.~\ref{Fig:NumExa_3}-\ref{Fig:NumExa_7} are shown in Fig.~\ref{Fig:SLP}. The service demand share and load pattern of the canonical domain (Fig.~\ref{Fig:NumExa_3}) are shown in Figs.~\ref{Fig:SLP1}~and~\ref{Fig:SLP6}, respectively. As indicated earlier, the conditions of the canonical domain (uniform spatial service demand distribution, uniform service demand share, and same amount of received interference per cell), result in a \textit{flat} load pattern, i.e., $\alpha_l=\alpha_k,\,\forall\,l\neq k$. In the example, each one of the 30 cells has a share of $3.33\%$ of the service demand, which  always results in a load equal to $0.91$. 

The service demand share of the topologies $\mathcal{T}_{\rm 1m}^{\prime}$ and $\mathcal{T}_{\rm 2m}^{\prime}$ created by the non-conformal mapping are shown in Figs.~\ref{Fig:SLP2} and \ref{Fig:SLP4}, respectively. In these cases, there is a small unbalance in the service demand share due to the fact that the mappings are only used to map the access points (and not the boundaries) from $\mathcal{R}$ to $\mathcal{R}^{\prime}$. Once in $\mathcal{R}^{\prime}$, cells are defined using standard Voronoi diagrams, i.e., each point is associated to its closest access point\footnote{Two additional methods, along with their pros and cons, are explained in \cite[V.C]{7399421}.}. A perfect service demand share could be attained, for instance, by using Algorithm~\ref{Alg:IterSitesLoc} to adjust the cells sizes of these topologies; however, they are intentionally kept in this manner to show that the power optimization proposed in Section~\ref{Sec:PO} does not require uniform service demand share. Their resulting load patterns, when uniform power allocation (cells transmit in data channels with the same power) is assumed, are shown in Figs.~\ref{Fig:SLP7}~and~\ref{Fig:SLP9}. Note that, as expected, uniform power allocation is not a good idea for topologies with very different access points densities, such as $\mathcal{T}_{\rm 1m}$, due to the high interference that is created. In Fig.~\ref{Fig:SLP7}, most of the cells get load factors greater than one (which in practice means outage), indicated by red circles.  In Fig.~\ref{Fig:SLP9} only one cell has load larger than one, but still the load pattern is very irregular, i.e., the network is far from being uniformly loaded.

The topologies obtained using Algorithm~\ref{Alg:IterSitesLoc} feature uniform service demand share as it can be seen from Figs.~\ref{Fig:SLP3} and \ref{Fig:SLP5}. The corresponding load patterns (under uniform power allocation) are shown in Figs.~\ref{Fig:SLP8} and \ref{Fig:SLP10}. In the light of these examples, it becomes clear that the more irregular the topology is, the less feasible the uniform power allocation assumption, and hence, the larger the need for power optimization. However, the average network load obviously depend on the service demand volume~($V$), but it is important to recall that this dependency is highly non-linear. Note that, while in case of $\mathcal{T}_{\rm 1c}$ (Fig.~\ref{Fig:SLP8}), there is one cell in outage and other three with loads very close to one, in case of $\mathcal{T}_{\rm 2c}$~(Fig.~\ref{Fig:SLP10}), cells are operating with loads below $0.6$. Evidently, the assumption of uniform power allocation is much less valid for $\mathcal{T}_{\rm 1c}$ than for $\mathcal{T}_{\rm 2c}$, as the former is much more \textit{irregular} than the latter, see Figs.~\ref{Fig:NumExa_5}~and~\ref{Fig:NumExa_7}. The following examples illustrates the use of power optimization.

\subsection{Load balancing through power optimization}\label{Sec:NR:LoadBalancingPO}
\begin{figure*}[t]
	   \centering	  					
			\subfloat[Optimized power for $\mathcal{T}_{\rm 2m}$.]
	    {\label{Fig:PO_1}
	    \includegraphics[width = 0.27\textwidth]{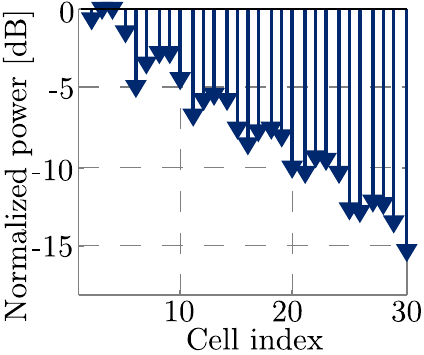}}	\hspace{0.3cm}	
							\subfloat[Spatial power allocation pattern for $\mathcal{T}_{\rm 2m}$.]
	    {\label{Fig:PO_2}
	    \includegraphics[width = 0.37\textwidth]{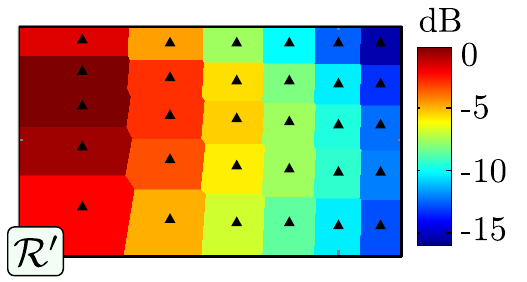}}	
			\subfloat[Uniform load pattern.]
	    {\label{Fig:PO_3}
	    \includegraphics[width = 0.27\textwidth]{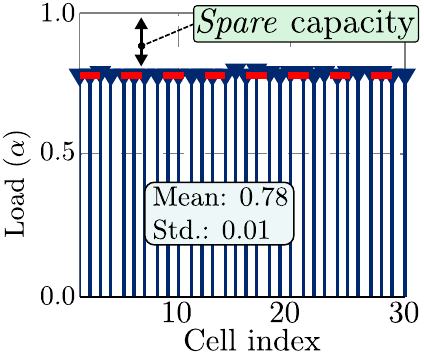}}	
\\[-0.2cm]
			\subfloat[Optimized power for $\mathcal{T}_{\rm 2c}$.]
	    {\label{Fig:PO_4}
	    \includegraphics[width = 0.27\textwidth]{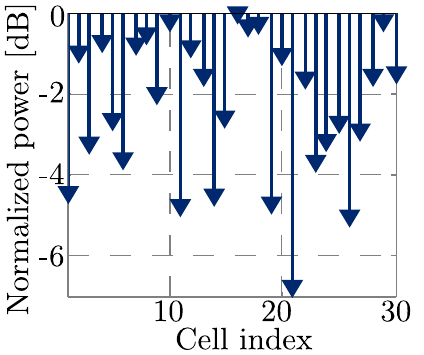}}\hspace{0.3cm}
							\subfloat[Spatial power allocation pattern for $\mathcal{T}_{\rm 2c}$.]
	    {\label{Fig:PO_5}
	    \includegraphics[width = 0.37\textwidth]{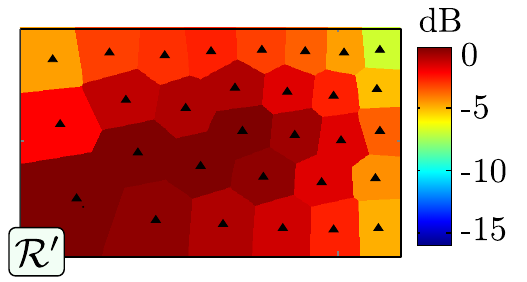}}	
			\subfloat[Uniform load pattern.]
	    {\label{Fig:PO_6}
	    \includegraphics[width = 0.27\textwidth]{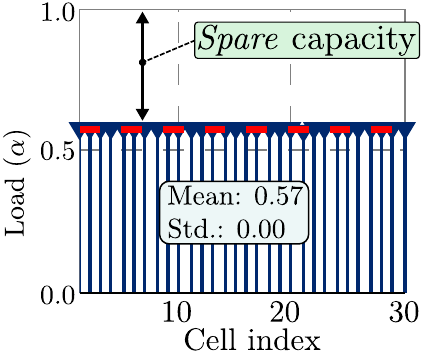}}	
	    \vspace{-0.3cm}\caption{Use of power optimization to obtain uniformly loaded networks with the highest \textit{spare} capacity.}
	    \label{Fig:PO}    		
\end{figure*}

As indicated before, optimizing the power allocated to the data channels is a convenient alternative to achieve load balancing, without need for transferring service demand from one cell to another; for instance, by adjusting the power of the very sensitive cell-specific reference signals~\cite{08:00004} that are used for cell-selection. In order to illustrate the use of power optimization, the topologies $\mathcal{T}_{\rm 2m}$ and~$\mathcal{T}_{\rm 2c}$ (Figs.~\ref{Fig:NumExa_6}~and~\ref{Fig:NumExa_7}), produced in $\mathcal{R}^{\prime}$ by the mappings $g_{\rm 2m}^{-1}$ and $g_{\rm 2c}^{-1}$, respectively, and the non-uniform spatial service demand distribution $\delta_2^{\prime}$ are considered. This does not imply any loss of generality with respect to its use in the physical domain $\mathcal{A}$, as the power optimization proposed in Section~\ref{Sec:PO} is topology and domain agnostic. 
Figs.~\ref{Fig:PO_1}~and~\ref{Fig:PO_4} show the optimized power vector $\mathbf{p}^{\star}$ in both cases. A visual representation is also provided in Figs.~\ref{Fig:PO_2}~and~\ref{Fig:PO_5}, where the power allocated to each cell is normalized and expressed in dB. Both figures have the same scale: from $0\,{\rm dB}$ (the highest power) to $-16\,{\rm dB}$. It can be observed that there is a certain correlation between the power allocated to each cell and its size, i.e., smaller cells tend to be allocated with less power. However, this is not a rigid rule as shown in the figures, but a trend that is just intuitively expected. The actual optimal power allocation depends on the spatial service demand distribution and network topology/geometry (site locations and cells). The resulting uniform load patterns are shown in Figs.~\ref{Fig:PO_3}~and~\ref{Fig:PO_6}, where it is evident that the proposed power optimization succeeds in finding a power allocation able to homogenize the load pattern in both cases.   
Thus, finding the solution of~\eqref{OP2} not only leads to a power allocation in which the cells are  uniformly loaded (a nice feature from network planning perspective), but also maximizes the \textit{spare} capacity in the network. Indeed, the resulting uniform load level $\bar{\alpha}$ is the minimum possible for a given $V$, $R_{\rm min}$, and $B$; in Appendix~\ref{App:CU}, uniqueness aspects are discussed. This is important because this \textit{spare} capacity, indicated in Figs.~\ref{Fig:PO_3}~and~\ref{Fig:PO_6}, provides a natural protection against instantaneous service demand variations. Recall that load values calculated using (\ref{Eq:baseline0}) are average figures.

\subsection{Final remarks}\label{Sec:NR:Comparative}
All in all, the results have shown the effectiveness and usability of both spatial mappings and centroidal-based algorithms with power Voronoi diagrams (Algorithm~\ref{Alg:IterSitesLoc}). As in many other instances, different approaches have advantages and drawbacks, and in this context the word `better' is not really suitable.  The use of one or another depends on preferences, requirements, available resources, and the characteristics of the particular problem to be addressed. A short list of practical \textsl{rules-of-thumbs} are provided next:
\begin{itemize}[leftmargin=0.75cm]
	\renewcommand{\labelitemi}{$\checkmark$} 
	\item \textbf{Scale of the problem}. Based on our experience, if the problem requires the deployment of a large number of cells (e.g., $L\in[75,1000]$), operating with spatial mappings is most like a good choice, because the execution of centroidal Voronoi algorithms could be very expensive, as centroids need to be computed for each cell at each iteration. In these cases, the \textit{compactness} of spatial mappings is really a desirable feature, as it was shown in the examples provided in~\cite{7399421}. For relatively small deployments ($L<75$), obtaining perfect service demand share is a nice \textit{plus} that can be obtained by means of Algorithm~\ref{Alg:IterSitesLoc}. However, as it was shown, uniform service demand share is not a requirement for power optimization.     
	\item \textbf{Context variables/assumptions}. The spatial service demand distributions used herein have been selected to be functions leading to closed-form mappings and centroid solutions. They also exemplify statistical independence and non-statistical independence. As indicated in~\cite{7399421}, conformal mapping as well as the mappings introduced herein not only admit, but require numerical solution in most of the  practical problems. Examples allowing analytic solutions, such as $\delta_1^{\prime}$ and $\delta_2^{\prime}$, are essentially reserved for theoretic/academic purposes. Therefore, more complex functions used to approximate the spatial service demand distribution would require numerical evaluation, same as spatial distributions completely given in numerical terms.  The particular structure of the $\delta$ that is assumed could be more suitable for one approach or another. It is also important to look at the definition of the physical domain. While conformal mapping provides a general setting for mapping arbitrary polygons onto rectangles, the assumption of defining~$\mathcal{A}$ as a rectangle would be valid in many practical contexts as well.  
	\item \textbf{Network fine-tuning}. The proposed framework is a tool for planning and optimization purposes; and it is a complementary approach to existing methods, such as system level simulations or stochastic geometry. Hence, it is perfectly valid to fine-tune the resulting/obtained network topologies using the methods proposed herein, i.e., adjusting cells coverage by means of power Voronoi diagrams or power optimization for load balancing as proposed in Section~\ref{Sec:PO}; or to resort to other alternatives, such as the power optimization for energy efficiency proposed in~\cite{6887352} or existing load balancing methods~\cite{6812287}. Planning and radio access optimization is a difficult problem, and hence, there is not a unique recipe. The  framework presented herein provides additional effective tools to aid at these tasks.

\end{itemize}

 		\section{Conclusions and Research Directions}\label{Sec:Conclusions}
Planning and optimization are tasks that certainly need to go \textit{hand-in-hand} to maximize the profit and performance of current and future cellular systems. In this work, key contributions to the planning and optimization framework based on canonical domains and spatial mappings, originally introduced in~\cite{7399421}, have been presented. These novelties include more general and versatile mappings, new algorithmic tools, and power optimization schemes. The results confirm the potential and promising research perspectives of the proposed framework, in which having an statistical description of the  spatial service demand distribution, is of utmost importance.  As it was explained, planning is a very tough problem, and to address it in the context of future 5G systems, several tools need to be combined to achieve the expected outcomes. In this sense, our framework is complementary to well-known existing methods, such as system level simulations or stochastic geometry, each of wish has advantages and drawbacks. However, the proposed framework provides not only another methodology (with pros and cons as well), but also a new angle to look at this problem. Because this idea is in its infancy, the authors are confident that many enhancements are to come and this contribution will benefit both academy and industry. 

 Our current research efforts are aligned in the following directions:
\begin{enumerate}
	\item \textsl{Indoor network planning}. The goal is to evolve the proposed framework for planning and optimization of indoor deployments, using cutting-edge mapping techniques developed by the authors, such as~\cite{hqr}.
	\item \textsl{Service demand in 3 dimensions}. Provide means to use the current art to study realistic cases where the service demand is given in 3 dimensions, e.g., to include buildings.
	\item \textsl{HetNets}: Heterogeneous networks is another direction to be addressed. Methods bases on the idea of composition and separation of tiers are currently under study. Also, integrating more general metrics to account with sectorization is another clear extension of this work.
	\item \textsl{Unmanned Aid Vehicles}~(UAVs): Due to its nature, the proposed framework is highly suitable to be used in the positioning of UAVs, where mobility can also be assumed for the access points. The temporal evolution of the service demand is another clear path to go.
\end{enumerate}
 		
 		% -- SECTION:ACKNOWLEDGEMENTS ---------------------------------	  	 	
		\section*{Acknowledgment}\label{Sec:Ack}
		\noindent This material is partly based upon works supported by the Academy of Finland under Grants 287249, 284811, and 284634.
	
		% Appendix
		\appendices
		\section{Power Optimization: Convergence and Uniqueness}\label{App:CU}

\begin{figure}[h]
	    		\centering	    	
	    		\includegraphics[width = 0.375\textwidth]{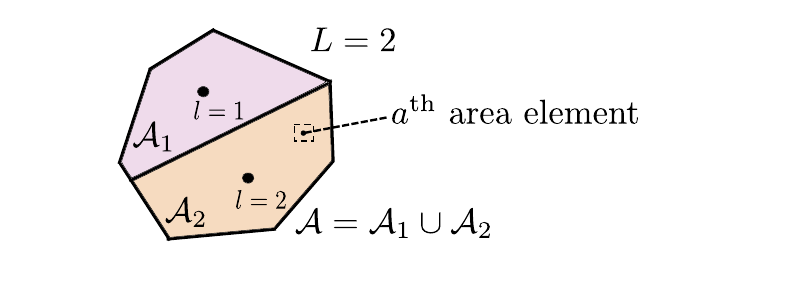}
	    		\caption{Toy example: small cellular network composed of two cells.}
	    		\label{Fig:TOY}    		
			\end{figure}

In this appendix, convergence and uniqueness aspects of the following minimization problem are discussed:
	\begin{eqnarray}
	\underset{\boldsymbol{p}}{\operatorname{minimize}}&&~~~{\rm Var}\{ \boldsymbol{\alpha} \},\label{OP2_a}\\
	{\rm subject~to:}&&\nonumber\\
	&&\boldsymbol{\alpha}=\boldsymbol{f}(\boldsymbol{\alpha}\,;\,\mathbf{p})\nonumber\\
	&&\mathbf{p}\in\mathbb{R}^L_{+},\nonumber
	\end{eqnarray}
where $\boldsymbol{\alpha}\in\mathbb{R}_{+}$, $\mathbf{p}\in\mathbb{R}_{+}$, and the $l^{\rm th}$ element of $\boldsymbol{\alpha}$, $\alpha_l$, is given by 
\begin{equation}
	\alpha_l= f_l(\boldsymbol{\alpha},\mathbf{p}) = K \sum_{a\in\mathcal{A}_l}\frac{\delta_a}{\log{\left(1+ \frac{p_{l}\,G_{l,a}}{\sum_{i=1,\,{i \neq l}}^{L}\, p_i\,G_{i,a}\,\alpha_i }\right)}}, \label{Eq:baseline0_a}
\end{equation}
with $\sum_{l=1}^{L}\sum_{a\in\mathcal{A}_l}\delta_a=1$, $K\in\mathbb{R}_{+}$, and $G_{l,a}\in[0,1)\,\,\forall\,\left(l\in\{1,2,\cdots,L\}\wedge a\in\mathcal{A}=\bigcup_{l=1}^{L}\mathcal{A}_l\right)$. The previous problem is a discrete version of~\eqref{OP2}, where the coverage region of each access point (i.e.,~the sets $\mathcal{A}_l$'s) is divided into many small area elements. For the sake of clarity, and without loss of generality, a basic network composed of two cells ($L=2$), as shown in Fig.~\ref{Fig:TOY}, is considered.
			
\subsection{Guaranteed Convergence of Interior Point Methods}

 The problem (\ref{OP2_a}) is quadratic.
 The gradients of the constraint functions \mbox{$\alpha_l - f_l(\boldsymbol{\alpha},\mathbf{p}) = 0$} and \mbox{$p_l > 0$} are linearly independent, 
 since by construction the matrix of gradients is diagonally dominant. Notice, that the diagonal elements are identically = 1.
 Thus, the convergence of interior point methods is guaranteed.

\subsection{Remark on Uniqueness of Minimal Load Vector}
Let us next focus on the discrete formulation of the simple configuration given above:
{\begin{eqnarray}
	\alpha_1= f_1(\boldsymbol{\alpha}, \mathbf{p}) = K \sum_{a\in\mathcal{A}_1}\frac{\delta_a}{\log{\left(1+\frac{1}{\alpha_2}\frac{p_1\,G_{1a}}{p_2\,G_{2a}}\right)}},\label{AA1}\\
	\alpha_2= f_2(\boldsymbol{\alpha}, \mathbf{p})=K \sum_{a\in\mathcal{A}_2}\frac{\delta_a}{\log{\left(1+\frac{1}{\alpha_1}\frac{p_2\,G_{2a}}{p_1\,G_{1a}}\right)}}.\label{BB1}
		\end{eqnarray}}	

Let us assume equilibrium and set $\bar{\alpha} = \alpha_1 = \alpha_2$ with naturally $ \bar{\alpha}\in\mathbb{R}_{+}$.
Next we consider a scaling $\bar{\alpha} \to c \bar{\alpha}$,
where $1/\bar{\alpha} > c > 0$ is a constant, 
in order to investigate whether for any value $c < 1$
there exists a power vector $\hat{\mathbf{p}}$ which also
minimizes the problem (\ref{OP2_a}).
Inserting the scaling into equations we get
{\begin{eqnarray}
	c \bar{\alpha}=K \sum_{a\in\mathcal{A}_1}\frac{c \delta_a}{\log{\left(1+\frac{1}{\bar{\alpha}}\frac{p_1\,G_{1a}}{p_2\,G_{2a}}\right)}},\label{AA2}\\
	c \bar{\alpha}=K\sum_{a\in\mathcal{A}_2}\frac{\delta_a}{\log{\left(1+\frac{1}{c\bar{\alpha}}\frac{p_2\,G_{2a}}{p_1\,G_{1a}}\right)}}.\label{BB2}
\end{eqnarray}}

Setting $p_1$ and $p_2$ to values obtained in the non-scaled
minimization in (\ref{AA2}) but rewriting (\ref{BB2}) as
\begin{equation}
	c \bar{\alpha}=K\sum_{a\in\mathcal{A}_2}\frac{\delta_a}{\log{\left(1+\frac{1}{c \bar{\alpha}}\frac{\hat{p}_2\,G_{2a}}{\hat{p}_1\,G_{1a}}\right)}},\label{BB3}
\end{equation}
we can search for a solution (minimizer) to an equation
(\ref{AA2}) = (\ref{BB3}), with $c$, $\hat{p}_1$, and
$\hat{p}_2$ as free parameters. 
This equation has only one fixed point with $c = 1$ and $p_1/p_2 = \hat{p}_1/\hat{p}_2$ and we conclude that
the obtained $\bar{\alpha}$ is optimal and unique up to scaling of the components of ${\mathbf{p}}$.

		% -- BIBLIOGRAPHY ------------------------------------------
	%	{\small
		%\bibliographystyle{ieeetr}
%\bibliography{TheBibliography,C:/Users/gonzald1/___WORK/___AVAILABLE_UTILITIES/LaTeXMyBibliography/BIB_StringDefinitions}}

		\end{document}